\documentclass[a4paper,conference]{IEEEtran} 
\ifCLASSINFOpdf
  \usepackage[pdftex,dvipdfm]{graphicx}
\else
\fi
\usepackage[cmex10]{amsmath}
\usepackage{amssymb,verbatim}
\usepackage[dvipdfm]{graphicx}
\begin{document}
%
\title{%
Design of  Quantum Stream Cipher: Part-I\\
-Lifting the Shannon Impossibility Theorem-  \\
}

\author{
\IEEEauthorblockN{Osamu HIROTA$^{1,2}$ \\}
\IEEEauthorblockA{
1. Quantum ICT Research Institute, Tamagawa University\\
6-1-1, Tamagawa-gakuen, Machida, Tokyo 194-8610, Japan\\
2. Research and Development Initiative, Chuo University, \\
1-13-27, Kasuga, Bunkyou-ku, Tokyo 112-8551, Japan\\
{\footnotesize\tt E-mail:hirota@lab.tamagawa.ac.jp} \vspace*{-2.64ex}}
}

\maketitle

\begin{abstract}
This paper is dedicated to the late Professor H.P.Yuen in commemoration to our 50-year friendship. 
He invented the concept of quantum stream cipher. 
It is designed based on a completely different concept from conventional ciphers. 
The purpose of this cipher is to provide information-theoretic security of long data and secret key 
with short key length. It is based on hiding the ciphertext of mathematical cipher with quantum noise, 
achieving unprecedented information-theoretic security in any cipher. 
The protocol corresponds to randomizing the ciphertext by means of differentiating the receiving performance of 
Bob with key and Eve without key according to the principle of quantum communication theory. 
In this paper, we introduce some progress on the specific method to develop from the standard type to 
generalized quantum stream cipher :Quantum Enigma Cipher. There are two methods for generalization. 
One is the additional randomization method by the product cipher form and the other is 
the M-th order extended quantum coding method. Here we discuss the former. 
The results proved that it has sufficient information-theoretic security against 
known plaintext attack on key in comparison with quantum data locking. 
The latter method will be reported in Part II.
\end{abstract}

%
\IEEEpeerreviewmaketitle

\section{Introduction}
The statistical communication theory by N.Wiener and C.E.Shannon started with a detailed analysis of the noise that affects
 communication performance. 
The noises for communications in an environment subject to classical physics were studied by Gibbs-Wiener modeling 
to statistical physics and their properties were formulated and systematized as noise theory by S. O. Rice [1] and others. 
Then, the optimal design theory of communication systems to cope with the noise was compiled by D.Midlleton, Y.W.Lee,
  H.L.van Trees, W.B.Davenport, J.M. Wozencraft, R.Gallager, R.S.Kennedy, J.H.Shapiro  et al.
It is so called the modern statistical communication theory [2$\sim$7].

On the other hand, the genesis of quantum information science began in the late 1960s with research into quantum communication theory. 
The noises for communications in an environment governed by quantum physics is called quantum noise.
It was developed as the optimal theory to deal with quantum noise.
This research aimed to predict the ultimate performance of optical communications. 

Early on, the study on the quantum effect to optical commuications started based on phenomenological formulations by D.Gabor [8], 
J.P.Gordon [9], H.A.Haus and others. C.W.Helstrom proposed the framework of his statistical quantum communication theory 
for quantum noise in 1967 based on Hilbert space theory, and published the book in 1976 [10]. 
A.S.Holevo proceeded the construction of the mathematical foundation for the statistical structure of quantum mechanics 
related to quantum communication [11].
 For further development of the quantum communication theory, 
as in the development of modern communication theory, 
the formulation was studied in the field of the mathematical information sciences.

For example, in the Soviet Union, by  A.S.Holevo, V.P.Belavkin and others, in the United States, 
by H.P.Yuen, J.H.Shapiro, J.Liu, S.D.Personic, V.W.Chan, S.Dolinar  and others under the guidance of R.S.Kennedy,
 and in Japan, by O.Hirota and others under the guidance of S.Ikehara, 
a global formulation for applicability to many fields was started in 1970s.

In the second stage, apart from Shannon information, the physics community began to build a quantum information theory 
that defines quantum information as information about quantum states and systematizes its information processing.
The former is a theory that will guide the future of current optical communications (Fig.1), while the latter aims to discuss 
the possibilities of quantum computers,and  microscopic phenomena as the physics.

The science and technology of communication must guarantee new capabilities without degrading current communication capabilities
 even if the issue is in quantum world. In particular, it is necessary to ensure $\lceil$ high speed, high efficiency, low cost,
 and high affinity with current network$\rfloor$.

These attempts were developed mainly at MIT's RLE (Fig.2) and Universisty of California at San Diego (Fig.3).  
On the other hand, in 1990s, Holevo and Belavkin  further developed the mathematical framework of quantum communication theory 
toward capacity formula and quantum stochastics process (Fig.4).

Under such a situation, Yuen disclosed an idea of quantum stream cipher in 2000.
This is a concept that involves developing technology to protect communication data and encryption scheme information by differentiating 
signal reception capabilities based on communications theory. It is not a concept that has been used in the past of hiding data 
by devising new encryption protocols.
  
Roughly speaking, this new concept can be said to be a prescription for retaining the convenience of mathematical encryption
 while enhancing the security of that encryption capability through the use of quantum noise.
Therefore, the communication system or modulation scheme itself becomes the encryption function.
As a result, an eavesdropper is battling against physical ciphertext in which a mathematically encrypted ciphertext is hidden by noise.
We have previously introduced this new cryptographic concept in two papers with the same title [12,13]. 
In this paper, 
the section II and III show the introduction of the background of the quantum stream cipher. 
The section IV and V show the concrete structure of the standard and generalized quantum stream cipher by a new description method.
The section VI and VII discuss the means of additional randomizations. 
In the appendix, the relationship between \textbf{the quantum stream cipher} and \textbf{the quantum data locking} 
related to Shannon impossible theorem is described.

\begin{figure}
\centering{\includegraphics[width=8.5cm]{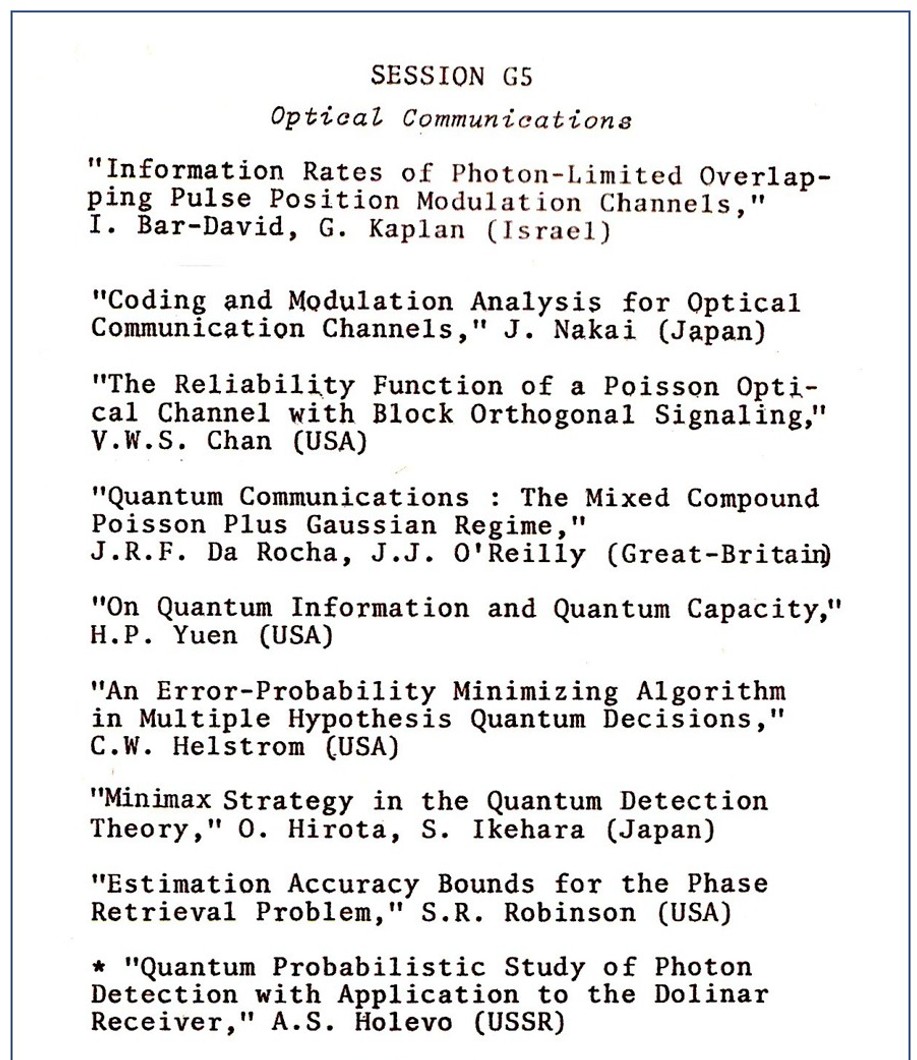}}
\caption{First meeting on quantum optical communication in IEEE International Symposium on Information Theory 1982 .
 The symposium was held in Les Arcs, France, and was co-chaired by Hellstrom and Picinbono.(by Hirota)}
\end{figure}

\begin{figure}
\centering{\includegraphics[width=6cm]{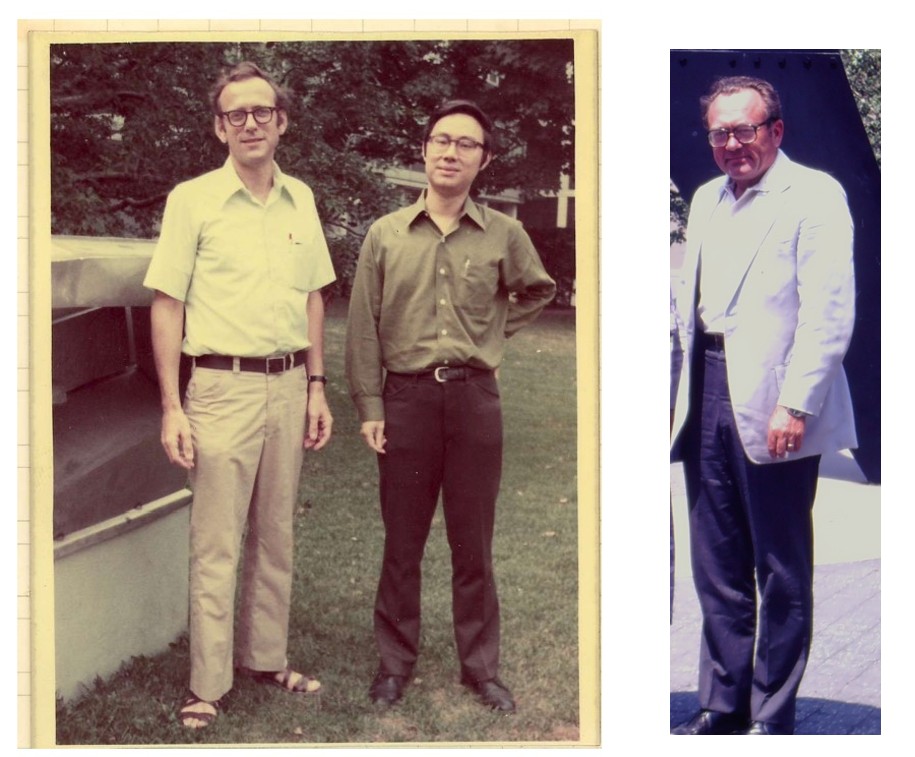}}
\caption{Left: R.S.Kennedy and H.P.Yuen at MIT, 1975 (by Hirota). Right: H.A.Haus at MIT, 1981 (by Hirota) }
\end{figure}

\begin{figure}
\centering{\includegraphics[width=5cm]{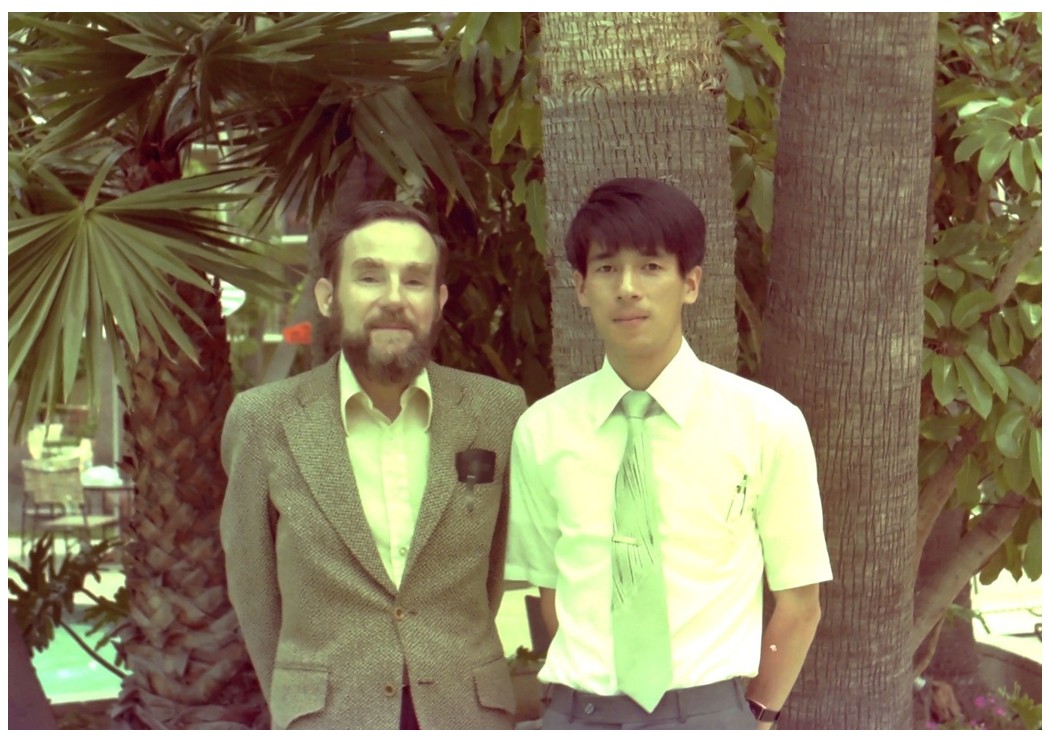}}
\caption{Left: C.W.Helstrom at University of California, San Diego, 1981 (by Hirota)}
\end{figure}

\begin{figure}
\centering{\includegraphics[width=6cm]{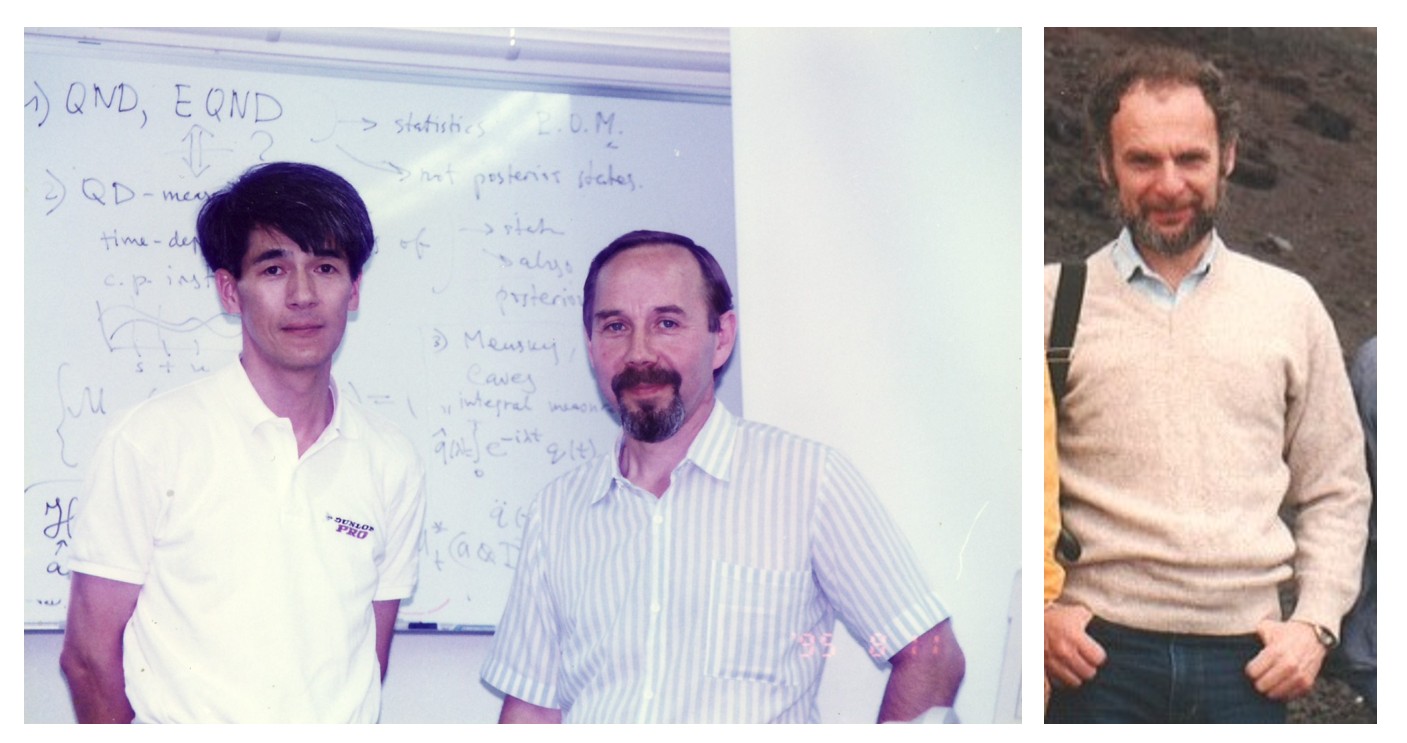}}
\caption{Left: A.S.Holevo (1996), Right: V.P.Belavkin (1998) at Tamagawa University, Tokyo (by Hirota)}
\end{figure}

\begin{figure}
\centering{\includegraphics[width=8cm]{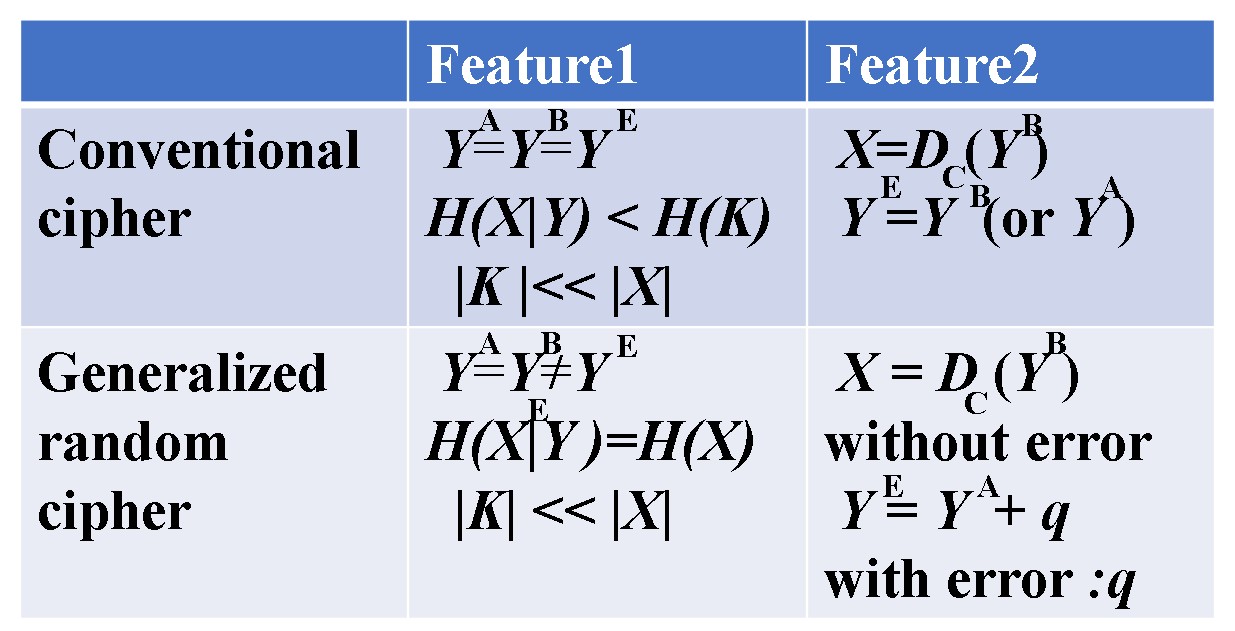}}
\caption{The conceptual differences between conventional mathematical cipher and generalized random cipher. 
$X$ is plaintext as natural sentence, $Y$ is ciphertext, $K$ is shared key, $|K|$ is key length, $D_C$ is decryption. 
A, B, E are indexes for Alice, Bob, and Eve.
Here, the following abbreviations are used:$Y^{E_q}=Y^E$ which is Eve's ciphertext with error. }
\end{figure}

\section{Concept of generalized random cipher in field of symmetric key cipher}

Before discussing quantum stream cipher, Yuen presented a hypothetical concept called generalized random cipher. 
Here, we give a brief explanation of the concept.
In general, the discussion on the information theoretic security of the symmetric key cipher is discussed based on 
Shannon-Massey random cipher. But due to \textbf{the Shannon impossible theorem} :$H(X|Y) \le H(K)$, the current technology cannot provide
 the information theoretic security against known plaintext attack. The detailed is given in [13].
To overcome it, Yuen proposed the concept of generalized random cipher explained in Fig.5. The important features are as follows:

Let us assume that the ciphertext received by a legitimate receiver (Bob) and the ciphertext received by an eavesdropper(Eve)are different, 
then the following situation is possible.
\begin{eqnarray}
&&H(Y^B_n|K,X_n)=0 \\
&&H(Y^{E_q}_n|K,X_n) \ne 0
\end{eqnarray}
where $Y^B_n,Y_n^{E_q}$ are ciphertext for Bob and Eve, respectively, $K$ means secret key
 and $X_n$ is plaintext.
Such ciphers are called the generalized random cipher. 
If such a situation could be implemented, it would be possible to realize a symmetric key cipher 
that has information-theoretically secure against known plaintext attacks even using a PRNG with a short key.
Furthermore, the possibility appears such that Eve cannot obtain the correct plaintext even with the correct key after communication. 
These features mean that the Shannon impopossibility theorem in the cryptology can be lifted. 
But, our goal is to improve the security against KPA on keys by lifting the theorem.

In general, the performance of the information theoretic security of the symmetric key ciphers can be evaluated 
by the spurious key or unicity distance. 
Thus, we adopt the unicity distance theory for our discussion to evaluate our proposed quantum stream cipher schemes. 
But we need a generalization of the unicity distance to evaluate a security of system in the class of the generalized random cipher.
The detailed discussions on the generalized unicity distances for such a cryptographic mechanism have been given in  [13].
The following is the short summary for those discussion.\\

\subsubsection{Ciphertext only attack}
The unicity distance of a ciphertext only attack on key is defined for the eavesdropper's ciphertext as follows:\\

${\bf Definition}1$: \\
Let $n^Q_0$ be the minimum length of the ciphertext that has zero key ambiguity for the eavesdropper's ciphertext.
Then it is given by
\begin{equation}
n^Q_0: H(K|Y^{Eq}_{n^Q_0})=0
\end{equation}
 $n^Q_0$ is called the unicity distance of ciphertext only attack for generalized random cipher.

Unlike the conventional type, the above equation does not depend on the statistical structure of the plaintext, but on the randomness of 
the ciphertext that can be obtained by the eavesdropper.\\

\subsubsection{Known plaintext attack}
The conventional random ciphers can achieve a large unicity distance for a ciphertext-only attack.
However, it is impossible to guarantee information-theoretic security more than a key length in the known plaintext attack.

Here, we consider a known-plaintext attack on generalized random cipher. First, the information-theoretic security evaluation 
for the known-plaintext attack is given as follows.\\

${\bf Definition}2$:\\
The unicity distance of known plaintext attacks for generalized random cipher is defined as follows:
\begin{equation}
n^Q_1: H(K|X_{n^Q_1}, Y^{Eq}_{n^Q_1})=0
\end{equation}

In the case of generalized random cipher, at least, the following can be expected.
\begin{equation}
|K| \ll n^Q_1 \le 2^{|K|}
\end{equation}
where $|K|$ is the key length. This performance is the most important in the practical applications.
This is not possible with existing cryptography theory.

The main objective of this paper is to explain a technique that quantitatively guarantees the information-theoretic security 
of known-plaintext attacks against the keys defined above.
In this case, the communication performance must be at least 100 Gbps for a communication distance of 1000 km, 
which is the performance of existing optical communication, because the symmetric key cipher is an encryption technology 
for high-speed communication.

To achieve this, it is convenient to use the quantum noise effects of light.
 As an example, if the signal system consists of non-orthogonal quantum states, an ideal random number error effect will automatically appear 
 in the received signal.
This is due to quantum irregularities (Born effect) when quantum superpositions are collapsed by measurement.
The detailed discussions will be given in the subsequent sections.\\

\section{Basic principles of quantum stream cipher from a communication theory perspective}

The security of the generalized random cipher relies on errors in the ciphertext of the mathematical cipher that an eavesdropper can obtain.
Our problem is how to realize such a mechanism. So we have following remark.\\

${\bf Remark} 1:$\\
Quantum stream cipher (Quantum noise randomized stream cipher) is a method for realizing generalized random cipher using the principles of 
statistical communication theory to create differences in errors between the ciphertext (or plaintext) received by a legitimate recipient (Bob)
 with key and the ciphertext that can be received by an eavesdropper (Eve) without key.
This differentiation is called advantage creation.
 Yuen named this concept \textbf{KCQ} (Keyed communication in quantum noise). \\

The above concept can be described as shown in Fig.6. Thus, we can expect that the quantum communication theory plays
 an essential role to create the differentiation to the error performance of Bob and Eve.
In the following, we will give a survey on a formulation for the error analysis of the ciphertexts (or plaintext) of legitimate receivers
 and eavesdroppers based on the quantum communication theory. 

\begin{figure}
\centering{\includegraphics[width=7cm]{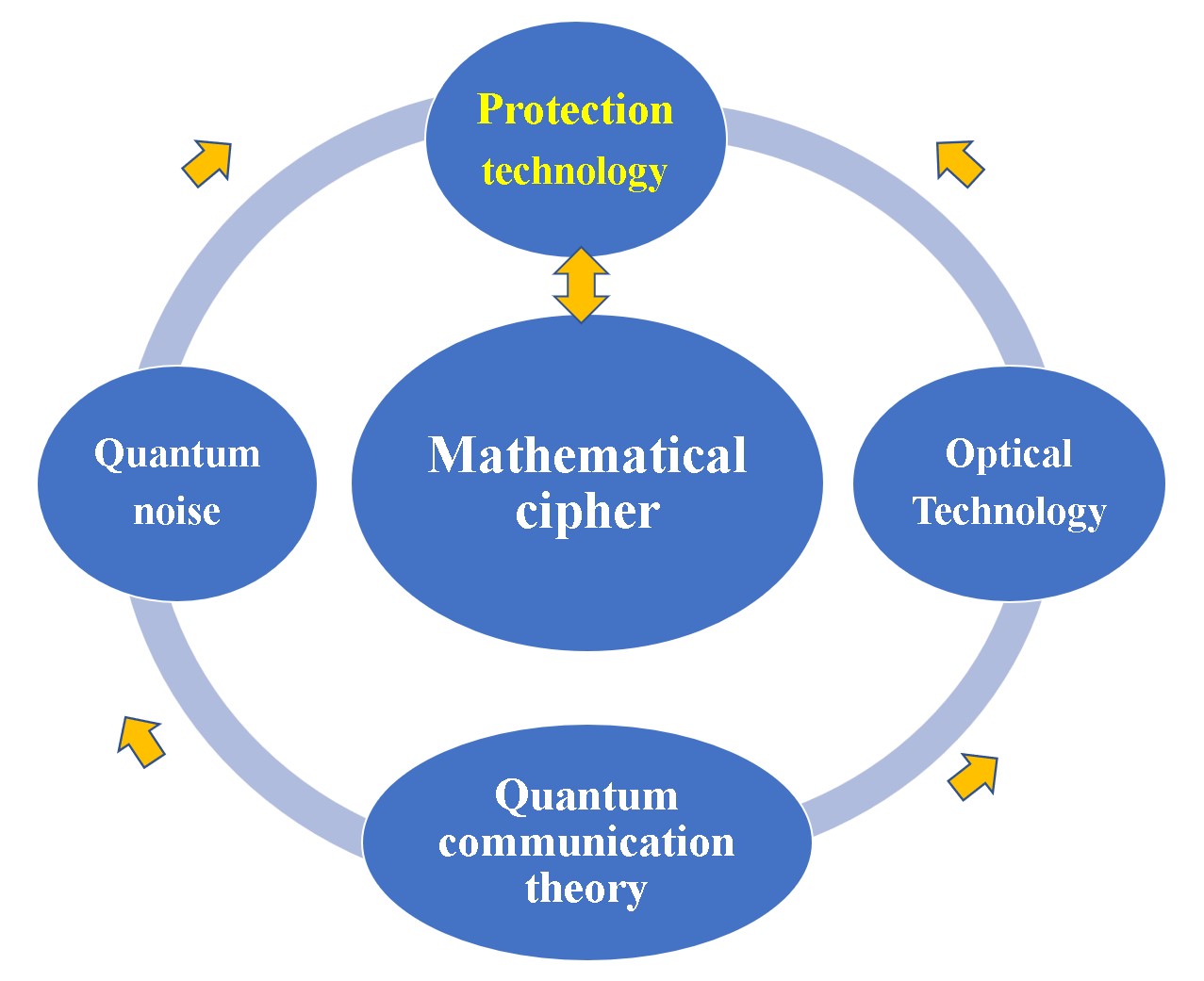}}
\caption{The concept of quantum stream cipher. The quantum stream cipher protocol corresponds to the protection technology of
 mathematical cipher. 
The ciphertext is protected by quantum noise effect designed by quantum communication theory}
\end{figure}

\subsection{Formulae of quantum communication theory}
\subsubsection{Generalized quantum measurement and decision operator}

The mathematical structure of quantum mechanics is given based on the Hilbert space theory constructed by von Neumann. 
The representation of the quantum measurement process is formulated using eigen state and eigen value of a self adjoint operator.
That is, it is defined as follows.\\

${\bf Definition} 3:$
Let us assume that the quantum system has the self adjoint operator as physical observable  ${\bf T}$ and its quantum state $|\Psi >$. 
The standard quantum measurement of the observable is described as follow:
\begin{equation}
{\bf T}|x >=x|x >, P(x) = Tr |\Psi ><\Psi | x >< x |
\end{equation}
where $| x >< x |$ is projection valued measure, and the above probability comes from the Born rule.

However, P. Benioff and others have been discussing the mathematical generalization of the projection process to describe 
the diversity of measurement processes. Yuen and Holevo have begun efforts to integrate this trend into communication theory.

Let us describe the formulation of quantum detection theory based on the above concept [10,14,15].
The set of density operator $\rho$ is a convex set, its extreme points being the one-dimensional projection. 
The corresponding states are called pure.
Any measurement with values in the real number space $R_M$ is described by an affine map of the set of the states (density operator)
 into the set of probability distributions on $R_M$.
Let us consider a generalized resolution of identity $\{{\bf X}_r ; r \in R_M \}$, i.e., the collections of Hermitean operators, 
stasfying ${\bf X}_r \ge 0, \sum_{r\in R_M}{\bf X}_r =I$.
$Tr \rho {\bf X}_r$ establishes the one-to-one correspondence between affine maps of the set of density operators into
 the set of probability distributions on $R_M$ and the resolution of identities.
In some cases, ${\bf X}_r$ is called the positive operator valued measure (POVM).  
In the quantum communication theory, the POVM is used as decision operator : $\{\Pi_m\}$ as follows:\\

${\bf Remark}2. :$
The decision operator does not just represent the probability of a quantum measurement process, but the probability of 
error or detection by the receiver's decision.
 In other words, it must be understood that it involves an operation by observers.\\
 
 On the above remark, from the origins of Helstrom's formulation, it is easier to understand if we interpret 
 the quantum decision operator as the generalization of  the Wald's decision function in the classical system [10].

In the classical communication theory, the decision is applied to the given probability function to variable of received signal. 
When one applies the standard quantum measurement, the probability function is given and the decision function is 
applied to its probability function same as the classical detection theory.
However, in quantum case, the decision operator does not need the explicit probability function of the variable 
of the received signal.
The decision operator directly outputs the probability value of the correctness or incorrectness of the decision 
without going through the probability function of the measurement process.
As an effect of the above fact, the possibility arises that the discrimination capability  to quantum signal
in the quantum measurement process will exceed the discrimination capability based on standard quantum measurement 
and its probability function. 
However, it can enjoy only in the discrete signal set.
 Its effect disappears in the case of continuous variable. That is, in the quantum estimation theory, 
there is no such quantum advantage. A reason has been discussed in [16] \\
\\
\subsubsection{Structure of quantum detection theory}
When one of $N$-ary quantum state signals $|\psi_m >=|\alpha_m>$ is received at each slot, the optimizing variable
 of the quantum measurement channel is described by a compact set of the decision operator: 
$\Pi_m$, $m=1,2,3, \dots ,N$.
Then these operations are interpreted as the projector acting on the quantum state of each slot, 
and these provide error or detection probabilities as follows:
\begin{eqnarray}
P(\alpha_l|\alpha_m)&=&Tr \rho_m \Pi_l, \quad m,l=1,2,3,\dots, N\nonumber \\
\rho_m&=&|\alpha_m ><\alpha_m|\nonumber \\
\sum_{l} \Pi_l &=&I, \quad \Pi_l \ge 0 \quad \forall l
\end{eqnarray}
where $I$ is the identity operator.
These are called quantum risk function in the detection theory.

The appearance of quantum effects in measurement process of signals and the result of the decision are simultaneously characterized 
by the above formula. 
Thus, the quantum Bayes rule is formulated without going through the likelihood ratio as follows:
\begin{equation}
{\bar {P}_e}=\min_{\{\Pi \}}\{1-\sum_{m=1}^{N} \xi_m Tr \rho_m \Pi_m\}
\end{equation}
where a priori probability must be $(\xi_m >0, \quad \forall m)$ for the admissibility in the decision theory.
The necessary and sufficient condition for $\{\Pi_m\}$ are given by Holevo [14] and Yuen [15]:\\

${\bf Theorem} 1$ $\{Holevo, Yuen\}$:
The necessary and sufficient condition for $\{\Pi_m\}$ on the quantum Bayes rule is given by 
\begin{eqnarray}
&&\Pi_m[\xi_m \rho_m -\xi_l \rho_l]\Pi_l =0, \quad \forall l,m \nonumber \\
&&\gamma -\xi_l \rho_l \ge 0, \quad \forall l \nonumber \\
&&\gamma =\sum_l \xi_l \rho_l \Pi_l
\end{eqnarray}

On the other hand, the quantum minimax rule for the non-trivial compact signal set is formulated by
\begin{equation}
{\bar {P}_e}=\max_{\{\xi\}} \min_{\{\Pi\}}\{1-\sum_{m=1}^{N} \xi_m Tr \rho_m \Pi_m\}
\end{equation}
and the necessary and sufficient conditions are as follows: [17]:\\

${\bf Theorem} 2$ $\{Hirota \cdot Ikehara\}$:
The necessary and sufficient condition for $\{\Pi_m\}$ on the quantum minimax rule is given by
\begin{eqnarray}
&&Tr \Pi_l \rho_l = Tr \Pi_m \rho_m, \quad \forall l,m \nonumber \\
&&\Pi_m[\xi_m \rho_m -\xi_l \rho_l]\Pi_l =0, \quad \forall l,m \nonumber \\
&&\gamma -\xi_l \rho_l \ge 0, \quad \forall l \nonumber \\
&&\gamma =\sum_l \xi_l \rho_l \Pi_l  
\end{eqnarray}

In general, it is very difficult to find the solutions of the above two quantum detection rules.
However, in the standard quantum stream cipher system, quantum state signals have a property of 
the covariant as defined below.[18,19]\\

${\bf Definition} 4$:\\ 
Let $G$ be a group with an operation $\circ$.
The set of quantum state signals is called group covarinat if there exist unitary operators $U_k(k\in G)$ such that
\begin{equation}
U_k|\psi_m >=|\psi_{k\circ m}>, \forall m,k \in G
\end{equation}
It characterizes quantum states $\{|\psi_m>, m\in G\}$.\\

The general properties of quantum Bayes rule for covariant case of multi parameters are given by Ban [20].
 One of the results for coherent state signals is as follows: \\

${\bf Theorem} 3$\\
If the signal set $\{|\alpha_m>\}$ is a covariant, the optimum POVM is given by using Gram operator $H$ as follows:
\begin{eqnarray}
&&\Pi_l =|\mu_l><\mu_l|  \\
&&|\mu_l>= H^{-1/2}|\alpha_l> \nonumber \\
&&H= \sum_{m=1}^M |\alpha_m><\alpha_m| \nonumber
\end{eqnarray}
and the optimum  quantum Bayes solution is 
\begin{equation}
{\bar P}_e =1-|<\alpha_1|H^{-1/2}|\alpha_1>|^2
\end{equation}
where $|\alpha_1>$ is the base state.\\

The error probability of Eve for $N$ covariant signals can be given as follows :
\begin{eqnarray}
{\bar P}_e^E &=&1- \frac{1}{(N)^2}(\sum_{m=1}^{N} {\sqrt \lambda_m})^2  \\
\lambda_m&=&\sum_{k=1}^{N} <\alpha_1|\alpha_k >u^{-(k-1)m} \nonumber
\end{eqnarray}
where $u=\exp[\pi i/M]$.
In addition,  Osaki showed that the worst priori probability in the quantum minimax rule for the covariant signals
 becomes the uniform distribution and the minimax solution is also given by Eq(13) and Eq(14) [16].\\

\subsection{Advantage creation by differentiation of quantum detection performance due to secret key}

Let us apply the above formulae to cryptanalysis.
Alice prepares $M$ sets of  the communication basis of two valued coherent states:
\begin{equation}
\{|\alpha_m >, |\alpha_m e^{i\pi}>\}, \quad m=1,2,\dots, M
\end{equation}
which corresponds to a binary data.
One of binary data is transmitted by one of communication basis which is randomly selected by PRNG.
Thus, the quantum  ciphertext is given by one of following $2M$ coherent states.
\begin{eqnarray}
&&\{|\alpha_1>, |\alpha_1 e^{i\Delta}>,  |\alpha_1 e^{i2\Delta}>, \dots,  |\alpha_1 e^{iM\Delta}>,\nonumber \\
&& |\alpha_{M+1}>, |\alpha_{M+1} e^{i\Delta+\pi}>,  |\alpha_{M+1} e^{i2\Delta+\pi}>,\nonumber \\
&& \dots,  |\alpha_{M+1}e^{iM\Delta+\pi}>\}
\end{eqnarray}

Bob can use the phase control device to convert back to binary state signals by using a running key 
from the same PRNG to the $2M$ state signals randomized by Alice's PRNG.
Then, the quantum detection model becomes the binary quantum states  independent of the communication basis.
Thus, the average error probability is given by Helstrom formula as follows [10]:
\begin{eqnarray}
{\bar {P}_e^B}&=&\min_{\{\Pi\}}\{1-\sum_{m=0}^{1} \xi_m Tr \rho_m^B \Pi_m \} \nonumber \\
&=&\frac{1}{2}[1- \sqrt {1-4\xi(1-\xi) Tr(\rho^B_0 \rho^B_1)}] \nonumber \\
&\ll& \frac{1}{2}
\end{eqnarray}

 On the other hand, ``in order for Eve to perform the cryptanalysis", she has to obtain the information of 
running key sequence by her quantum measurement to $2M$-ary quantum ciphertext.
The first step in the procedure leading to an attack is to receive a signal flowing through the real communication channel. 
The average minimum error probability for the adopted quantum state signal scheme
 (or equivalently maximum detection probability) can be given by the formulae: Eq(9), Eq(10),Eq (11). 
For $M \gg 1$, one can expect
\begin{equation}
{\bar {P}^E_e}=\max_{\{\xi\}} \min_{\{\Pi\}}\{1-\sum_{m=1}^{2M} \xi_m Tr \rho_m \Pi_m\} \sim 1-\frac{1}{M}
\end{equation}
This provides the evaluation of the theoretical accuracy of the ciphertext that the eavesdropper can obtain.

If Eve were to attempt to decode the binary data directly, she would adopt the binary quantum optimal measurement 
for the following mixed quantum states.
\begin{eqnarray}
\rho^E_0 &=&\frac{1}{M}\sum_{m=1}^{M} |\alpha_{(m=even)}><\alpha_{(m=even)}| \nonumber \\
\rho^E_1 &=&\frac{1}{M}\sum_{m=1}^{M} |\alpha_{(m=odd)}><\alpha_{(m=odd)}| 
\end{eqnarray}
This structure of mixed state is called doubly symmetric mixed state, and the quantum Bayes(also minimax) solution for 
 such as mixed states of coherent state was given in [21].
Then the average error probability for binary data  in the case of $M\gg 1$ is given as follows: 
\begin{equation}
{\bar {P}_e^E}=\max_{\{\xi\}}\min_{\{\Pi\}}\{1-\frac{1}{2}\sum_{l=0}^{1} Tr \rho_l^E \Pi_l \}\sim \frac{1}{2}
\end{equation}
The difference between Eq (18) vs Eq (19) or Eq (18) vs Eq (21) is called \textbf{the advantage creation by a secret key}, 
which is the concrete result from the keyed communication in quantum noise: KCQ [22].

\subsection{Quantitative evaluation of key security}
Quantitative evaluation of the security of quantum stream cipher may be carried out directly by the error characteristics
 of the eavesdropper's ciphertext,
 but the generalized unicity distance is cryptographically preferable to evaluate known-plaintext attacks. 
 It is given as follows [23]:\\

${\bf Theorem} 4$ $\{Yuen \cdot Nair\}$\\
The lower bound of the generalized unicity distance for KPA  is given as follows:
\begin{equation}
n^Q_1 \ge \frac{|K|}{C_1},\quad C_1=\max_{\{\Pi^E\}}I(K^R;Y^{E_q})
\end{equation}
where  $C_1$ is the maximum amount of accessible information by the eavesdropper's measurement from set of quantum states 
with running keys as information.
 
The theory of maximaization of mutual information is twinned with the above quantum decision theory and 
has the same theoretical structure as the optimal theory. It is derived by Holevo [14]:\\

${\bf Theorem} 5$\{Holevo\}\\
The necessary condition for maximum mutual information with respect to the decision operators for a simple set of states 
is given as follows:
\begin{eqnarray}
&&P(j|i) =Tr \rho_i \Pi_j \nonumber  \\
&& {\bf F}_j = \sum_{l} \xi_k \rho_k \log  \{\frac {P(j|l)}{\sum_k \xi_k P(j|k)}\} \nonumber \\
&& \Pi_j[{\bf F}_j - {\bf F}_i]\Pi_i =0,  \forall i,j 
\end{eqnarray}

Let us consider the concrete property of the above. Several properties of the mutual information is discussed by Ban and Osaki. 
The most convenient results are as follows [24]:\\

${\bf Theorem}6$$\{Ban \cdot Osaki \}$\\
When the signal states are the group covariant, the quantum Bayes and minimax decision operators satisfy 
the above necessary condition for the mutual information.\\

Now we still have a difficult problem that  is a proof of the sufficiency.
According to Osaki's numerical analysis, the quantum minimax decision operator may provide the maximum mutual information
 in the practical region such as $|\alpha|^2 \gg 1$ in the set of coherent states.

This requires a numerical analysis for the visualization. Fortunately, in the case of $M \gg 1$,
 a set of the quantum states can be treated as a quantum state system corresponding to a continuous signal. 
So the quantum optimum measurement can be approximated by Yuen-Lax quantum Cramer-Rao bound [25].\\

${\bf Theorem 7}$ : 
The estimation bound for complex amplitudes are given by following formula.
\begin{equation}
Var ({\hat \alpha}) \ge \frac{1}{Tr \rho L_R L_R^{\dagger} }
\end{equation}
where the right logarithm derivative $L_R$ is defined by 
\begin{equation}
\frac{\partial \rho}{\partial \alpha} = L_R^{\dagger} \rho 
\end{equation}
And its solution is as follows:
\begin{equation}
L_R = {\bf a}
\end{equation}
where ${\bf a}$ is a photon annihilation operator, and it corresponds to a heterodyne measurement.\\

Thus, we can adopt the mutual information based on the heterodyne receiver in the case of $M \gg 1$.
Apart from these mathematical difficulties, the method of approximating by an upper bound on the mutual information 
 is valid based on the following theorem [18,26,27].\\

${\bf Theorem} 8$\\
 $\{Holevo\cdot Schumacher \cdot Westmoreland \}$\\
The upper bound of maximum mutual information and the capacity are given by the Holevo information as follows:
\begin{eqnarray}
&&C_1 \le S( \rho_T) - \sum_{k=1}^{M} \xi_k S(\rho_k)=I_H\\
&&where \quad \rho_T = \sum_{k=1}^M \xi_k \rho_k, \nonumber\\
&&Then \quad C_H=\max_{\xi} I_H
\end{eqnarray}
where $S(\rho)$ is the von Neumann entropy.\\

From the above results, the optimization theory to quantum measurement and decision so called quantum communication theory 
allows for quantitative evaluation of the information theoretic security of the quantum stream cipher through the evaluation of $C_1$.
 
 Now our challenge is to find a way to achieve without the degradation of communication performance the following:
 \begin{equation}
 \min_{\{\Xi\}} C_1\longrightarrow 0
 \end{equation}
 where $\Xi$ is the randomization. In the following sections, we will show some examples of the randomization.\\

\section{Specific structure of protocol of standard quantum stream cipher:Y-00}
\subsection{Signal structure of basic model($\alpha\eta$)}

The basic model for the explanation of the principle is given in Fig.7.
Alice and Bob share a secret key $K$ and PRNG as a symmetric key cipher. The key length is $|K|= 100 \sim 1000$ bits. 
The key is extended by the PRNG. It is called a running key.
 The output bit sequence of the PRNG is divided by each $log M$ bits,
 and each $log M$ bits is used as the running key: $K^R:\{K^R_j=1,2,\dots, M\}$ for the selection of the communication basis
  which is a set of two coherent states:
$\{|\alpha e^{i\theta_j}\rangle, |\alpha e^{i(\theta_j+\pi)}\rangle \}$ in the PSK scheme [23]. 
That is, when a running key appears, one of communication basis corresponding to the running key is chosen.
 Then, the binary data $x\in X$ is transmitted by $|\alpha e^{i\theta_j}\rangle$ or 
 $ |\alpha e^{i(\theta_j+\pi)}\rangle $ of the basis. 
A mapping pattern from running keys to bases of two coherent states is given by the next relation
 in the basic model by the phase modulation:
\begin{equation}
{\cal{L}}=
\left(
\begin{array}{cccc}
K^R_j\\
{\bar \theta}_j 
\end {array}
\right ) 
=
\left(
\begin{array}{ccccc}
1 & 2 & 3 & \dots &M\\
 {\bar \theta}_1 &{\bar \theta}_2&{\bar \theta}_3& \dots&{\bar \theta}_M
\end {array}
\right ) 
\end{equation}
where the mapping $K^R_j \rightarrow {\bar \theta}_j$  means that 
$K^R_j$ determines one communication basis ${\bar \theta}_j=\{\theta_j, \theta_j + \pi \}$.

Thus, in the case of the basic model, the structure of phase signal is given by 
\begin{eqnarray}
\theta_m &=&f(K^R_j,X), \quad m=1,2,3,\dots 2M \\
&&K^R_j=1,2,3, \dots, M, \quad X=0,1 \nonumber
\end{eqnarray}
where $f(\cdot)$ is a mapping function from  digital signals to a physical signal by optical modulater.
The plaintext $X$= 0 and $X$=1 are regularly arranged between adjacent signals on the circumference on the phase space. 
So if the running key is even, the plaintext is 0, and if it is odd, it is 1. This corresponds to a geometrical 
relation on the phase space between dplaintext and communication basis. 
In the mathematical sense, the mapping function $f(\cdot)$ is injective w.r.t $K^R_j$ and $X$.

The quantum state sequence that transmits the information of $\theta_m$ is described as follows:
\begin{eqnarray}
|\Psi \rangle_T &=&|\alpha(K^{R}_j,X) \rangle_1 |\alpha(K^{R}_j,X) \rangle|\alpha(K^{R}_j,X) \rangle \dots \nonumber \\
&=&|\alpha_m \rangle_1 |\alpha_m \rangle_2 |\alpha_m \rangle_3 \dots
\end{eqnarray}
where $|\alpha_m \rangle$ is one of 2$M$ coherent states described by $\alpha_m=\alpha e^{i\theta_m}$; $m =1,2,3, \dots, 2M$.
So the coherent state as the signal is given by 
\begin{equation}
|\alpha_m>=|\alpha e^{i\theta_m}>
\end{equation}
Here, the phase difference between adjacent signals is $\pi /M$ or the signal distance is $\Delta_p \cong \pi|\alpha|/M$.
Consequently, the signal states satisfy the following relation under the fixed $|\alpha|$:
\begin{equation}
|\langle \alpha_m |\alpha_{m+1} \rangle|^2 \sim 1, \forall m, \quad M \gg 1
\end{equation}
This means that these two states cannot be discriminated due to the principle of the quantum detection theory.
Such an indistinguishable region $\Lambda$ is called the quantum noise masking region. 
When the signal power is strong, it is designed to be around 10.

\subsection{Receiver systems for Bob and Eve}

Bob has the information of selection of the communication basis. Hence, Bob can adopt the phase shift device 
before the quantum detection, and he can adaptively convert the signal detection problem to 
 2$M$-ary to the binary detection problem.
So he can use the Helstrom receiver to the binary coherent states as follows: 
\begin{equation}
\{|\psi_X> \}=\{|\alpha^{i\theta_m} >, |\alpha^{i(\theta_m +\pi)} >\}, \forall m
\end{equation}
If its received power were not so small, it could be replaced by homodyne receiver for the binaly signals.

On the other hand, since Eve does not have the information of the running key from PRNG, 
Eve has to adopt $2M$-ary quantum detection scheme for
\begin{equation}
\{|\psi_m> \} = \{|\alpha^{i\theta_m}>, m=1,2,\dots, 2M \} 
\end{equation}
because she cannot use the phase shift device (phase controller).
Thus, according to the principle of quantum statistical communication theory, differences in reception capabilities will emerge.
 As a result, Eve's receiver will be inferior to Bob's. 
That is, the error of Bob's receiver is small, and that of Eve is inferiority.
It is equivalent that the eavesdropper receives the ciphertext of the mathematical cipher based on PRNG with errors 
due to quantum noise. This is called advantage creation. Fig.7 shows the visualization of such processes.

\subsection{Overlap selection keying(OSK)}
The OSK is a mechanism to randomize the relation between the running key and plaintext(data) in the phase signal
 by a product cipher form [28].
That is, OSK provides an effect that the relation between data and given basis by flipping the data 0 and 1 based on 
a binary sequence from a branch of the  PRNG for the selection of communication basis is randomized.  
By this method, the geometrical relation on the phase space of data $X$ and communication basis $K^R$ is broken.
 In addition, the plaintext becomes from $X$ to $X_{Y00}=g(K^R_b,X)$, where $g(\cdot)$ is an encryption function. 
So the structure of the phase signal is replaced as follows:
\begin{eqnarray}
&&\theta_m =f(K^R_j, X_{Y00}), \quad m=1,2,3,\dots 2M \\
&&K^R_j=1,2,\dots M, \nonumber \\
&&X_{Y00}=g(K^R_b, X):\{0,1\}, K^R_b = 0,1 \nonumber 
\end{eqnarray}
 In this case, the function $f(\cdot)$ is not injective.
That is, when $\theta_m$ is given, even if $K^R_j$ is known, $X_{Y00}$ remains undetermined, or the opposite is true.
Although this mechanism looks like the structure of ``a product cipher", 
its role is quite different from that in traditional ciphers.

Now let us explain what happens as a result of the above.
Physically Eve can only measure the phase signal as follows: 
\begin{equation}
{\hat \theta}_m =f(K^R_j, X_{Y00})+ q\\
\end{equation}
where $q$ is decision error by quantum noise which is a true random number.
So she has to use the measurement values with error to proceed to cryptanalysis on $K^R_j$ for secret key analysis 
or on plaintext. 

If Eve were to directly perform the measurement of plaintexts $X_{Y00}$, the quantum state for Eve becomes 
from Eq(20) to the following.
\begin{equation}
\rho^E_0 = \rho^E_1 \longrightarrow \bar{P}^E = \frac{1}{2}
\end{equation}
Thus, it is impossible to discriminate such states.

If Eve adopts $2M$-ary detection, the effect of OSK becomes as follows (Fig.8): \\
The phase information becomes the single-valued real function with independent two variables: $K^R_j$ and $X_{Y00}$.
In general, these two variables cannot be uniquely determined from the measured phase.
When quantum noise is small, the error in the phase signal corresponds mainly to the error in $K^R_j$.
In the case without OSK, the information of the plaintext may be useful to analyze the structure of the cipher.
However, in the case with OSK, we can expect that it becomes impossible to estimate $K^R_j$ from the measured values, 
even if the true plaintext $X$ is known. The basic model including OSK is called standard quantum stream cipher, 
and its protocol is called the Y-00 protocol.
Futher detailes are given in the appendix A and B.

\begin{figure}
\centering{\includegraphics[width=8cm]{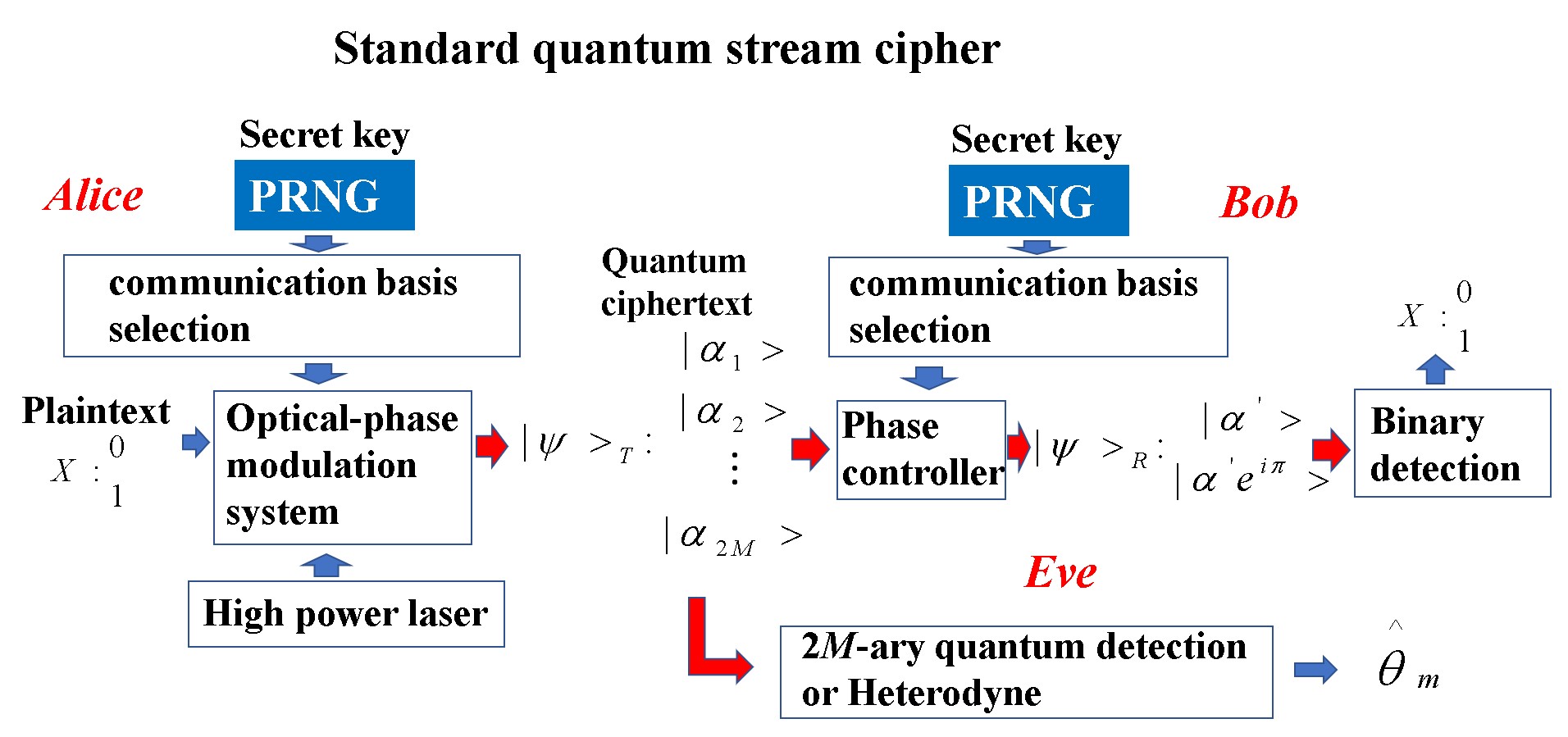}}
\caption{Structure of the basic model for the explanation of the concept. 
The optical modulator selects one  basis from a set of $M$ communication bases according to a running key. 
It is done by phase controller. Then it performs binary PSK by using that basis. When OSK is added, 
the plaintext is randomized by the binary sequence from PRNG. }
\end{figure}

\begin{figure}
\centering{\includegraphics[width=7cm]{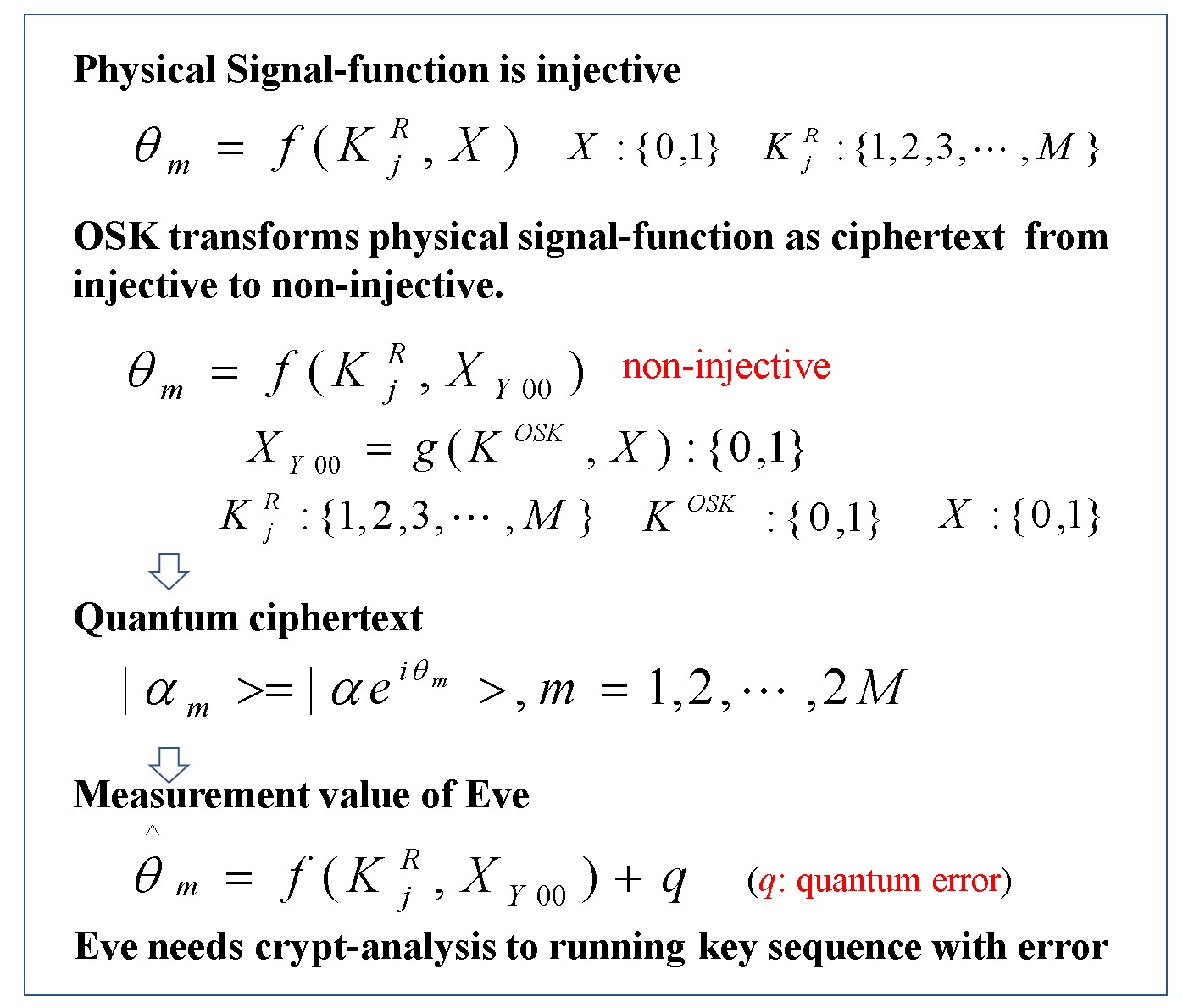}}
\caption{Effect of OSK to multi values detection. In the case of the basic model, physical ciphertext
 by modulation is injective w.r.t. running key and plaintext.
 In the case with OSK, it becomes non-injective w.r.t. plaintext. 
 In addition, Eve can only obtain the value of ${\hat \theta}_m$ with error. To binary detection, it is given by Eq(39).}
\end{figure}

\subsection{Cryptographic properties}
In the above system, according to Eq(39), a ciphertext only attack on data (plaintext) may have the same 
information theoretic security as a one time pad, despite the short secret key scheme. That is, $H(X|Y)=H(X)$.
 
When it comes to the information theoretic security of secret keys, the comparison discussion is a bit more complicated.
In mathematical cryptography, information theoretic security against ciphertext-only attack on key is guaranteed by 
Shannon-Massey random cipher with randomization of the plaintext.
 But it requires the exact prior information on statistics of data and the rate reduction by data randomization.
In addition, the information-theoretic security against known plaintext attacks on key is an order of the key length.

For quantum stream ciphers using OSK, known plaintext attacks are ineffective, so any plaintext can be guaranteed (see Appendix B). 
However, when quantum noise effects are small, the unicity distance  is not very large.

In conclusion, a distinctive feature of the standard quantum stream ciphers is that even short secret key ciphers can demonstrate
 performance equivalent to that of a one-time pad against ciphertext-only attacks on data (plaintext).
But, the quantum noise effect is not so large, and the security on key is limited.
It means that the advantage of the quantum stream cipher should be evaluated based on the key security.

Thus, the question is whether it is possible to significantly improve the information theoretic security of secret keys
 in the case of general plaintext (normal text) or known plaintexts.

\section{From standard to generalized quantum stream cipher}

Standard quantum stream ciphers, as Yuen himself explains, serve as an explanation of the principle.
He recommended researching its generalization based on standard quantum stream ciphers.
It is called generalized quantum stream cipher or Quantum Enigma Cipher (That is different from quantum enigma machine of Lloyd). 
In this section, we will show some examples. 
There are two methods for doing so, as shown below [13].

\subsection{Additional randomization method by product cipher form}
In generalized random ciphers of the section II, the ciphertext or running key sequence obtained by an eavesdropper
 must be perturbed by a true noise.
If the noise effect is small in the standard model, it naturally cannot provide sufficient 
information-theoretic security from the above theory.
For realistic applications, it is necessary to develop techniques to enhance the quantum noise effect.

The problem to make increasing noise effect, as described above, has little precedent in information theory and
 is the exact opposite of  what has been done so far.
That is, the technological development is to increase the noise effect of the eavesdropper, but it must be no effect 
the legitimate communicator.
In the channel model of receiving process of an eavesdropper, a research to enhance the error probability of the ciphertext of Eve
 or to reduce its maximum mutual information is called ``Randomization technique".
One of the example is Deliberate Signal Randomization (\textbf{DSR}) proposed by Yuen-Nair-Kumar[23]. 
Another example of a product cipher form is called ``Quantum noise diffusion mapping"(\textbf{QNDM}) [29].

\subsection{M-th order extended quantum code modulation}
\subsubsection{Coherent PPM}
There is a method to realize a generalized quantum stream cipher without using the above randomization technique.
This mechanism treats information as $M$-th extended code and configures  a transmitted signal system of $M$-ary PPM 
based on on-off keying. 
The $M$-ary PPM signals of Alice are then converted into  a pseudo-waveform signal of the $M$- ary slot by unitary transformations
driven by pseudo-random numbers generator.  It is called a coherent Pulse Position Modulation (CPPM).
This scheme is very effective in degrading the receiver performance of an eavesdropper and was discussed in detail in reference [22].

\subsubsection{Frequency-phase PPM}

The CPPM needs to expand  the baseband  to avoid the time delay for the encryption and description.
In order to avoid the drawback, we proposed the frequency-phase PPM scheme [30] based on the symplectic transformation 
for coherent state ensemble [31]. The detailed discussions on CPPM and frequency-phase PPM will be in the next paper.

\section{Quantum Noise-Diffusion Mapping (QNDM)}

In order to enhance the performance of the information theoretic security against KPA, we proposed 
the additional randomization technique so called ``Quantum noise diffusion mapping" in 2007 [29]. 
In this section, we discuss the propoerty from the viewpoint of unicity distance.

\subsection{Structure of QNDM and its effect}
\subsubsection{Structure of mapping}
A basic idea of the error enhancement techniques for a provable security has been described in the reference [23]
so called the deliberate signal randomization (DSR) which is a method without a shared key. 
Besides, the error performance of Bob is degraded, so one will need an appropriate design. 

\begin{figure}
\centering{\includegraphics[width=8cm]{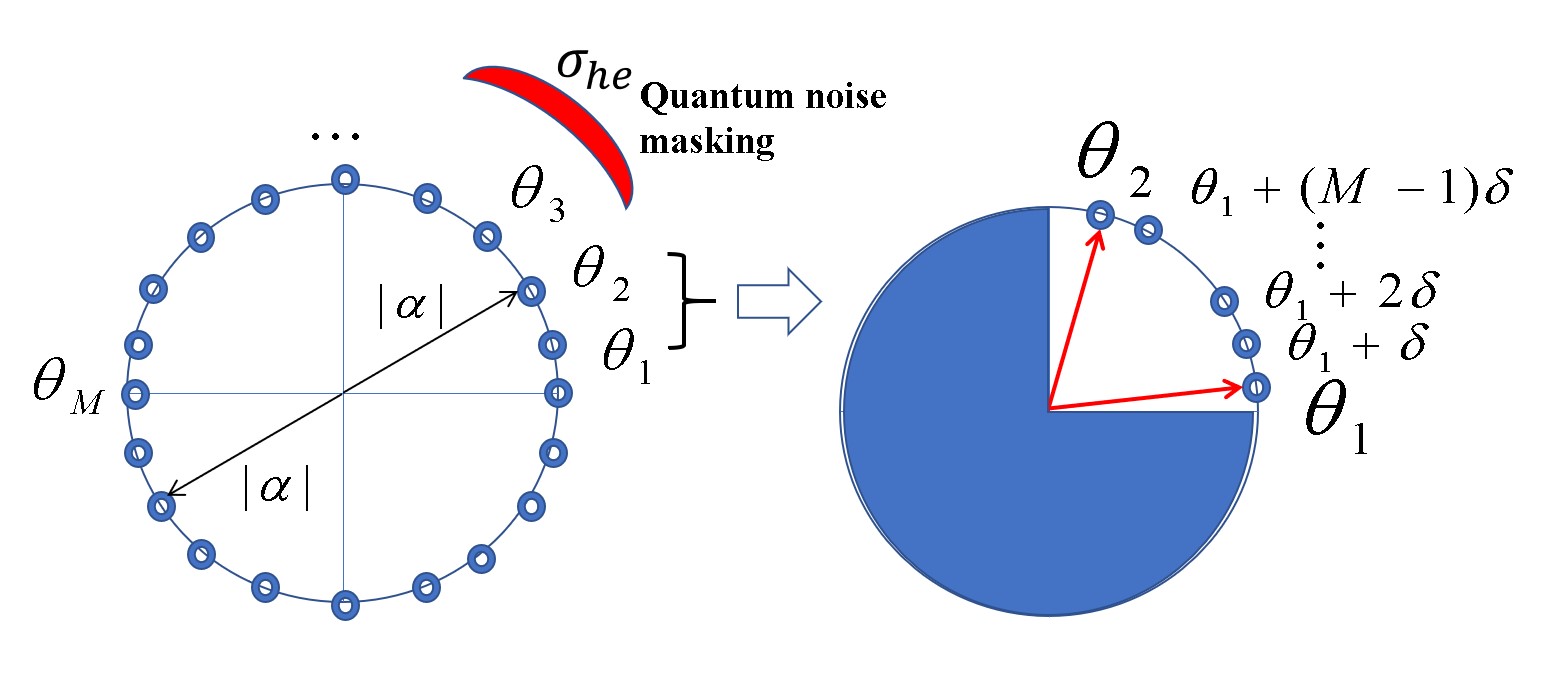}}
\caption{Signal constellation of the QNDM on phase space. 
The left is the first phase signal constellation: ${\cal L}_1$, and the right is enlarged view between $\theta_1$ and $\theta_2$.
There are $M$ signals between $\theta_j$ and $\theta_{j+1}$, so a total of $M^2$ signals are used.}
\end{figure}

As another method, let us here introduce keyed randomization such as QNDM [13,29] as mentioned in the previous section. 
 This is a randomization by an additional PRNG with an additional shared key, which is a kind of product cipher form.
 This randomization has such an advantage that it does not affect the error performance of Bob's receiver.  

Let us introduce the mechanism of this mapping scheme.
First, divide the circumference of the phase signal by $2M$.
Let the each region be $\Gamma_l, l=1,2,\dots, 2M$. It corresponds to $|\theta_{j+1} - \theta_j|$ which is the phase difference
 in the original mapping ${\cal L}_1$.

The first PRNG-1 choses  a mapping  pattern  from many mapping patterns
 ${\cal {L}}_1, {\cal {L}}_2,\dots,  {\cal {L}}_M$. However, each mapping pattern of the set $\{{\cal {L}}_k\}$ is designed as follows:
\begin{eqnarray}
{\cal{L}}_1&=&
\left(
\begin{array}{cccc}
K^R_j\\
{\bar \theta}_j 
\end {array}
\right ) 
=
\left(
\begin{array}{ccccc}
1 & 2 & 3 & \dots &M\\
 {\bar \theta}_1 &{\bar \theta}_2&{\bar \theta}_3& \dots&{\bar \theta}_M
\end {array}
\right ) \nonumber \\
{\cal{L}}_2&=&
\left(
\begin{array}{ccccc}
2 &  \dots &M  &1\\
 {\bar \theta}_1+\delta & \dots& {\bar \theta}_{M-1}+\delta  &  {\bar \theta}_M+\delta
\end {array}
\right )\nonumber \\
{\cal{L}}_3&=&
\left(
\begin{array}{ccccc}
3 &  \dots &1&2\\
{\bar \theta}_1+2\delta & \dots& {\bar \theta}_{M-1}+2\delta&{\bar \theta}_M+2\delta
\end {array}
\right )\nonumber \\
\vdots \\
{\cal{L}}_M&=&
\left(
\begin{array}{ccccc}
M & \dots &M-1 \\
{\bar \theta}_1+(M-1)\delta &\dots&{\bar \theta}_M+(M-1)\delta \nonumber
\end {array}
\right )
\end{eqnarray}
where ${\bar \theta}=\{\theta_j, \theta_j + \pi\}$ is a communication basis, and  
 $\delta = |\theta_{j+1} - \theta_j|/M$. It is the value that the phase difference between neighbors
 in the original standard scheme is divided by $M$. 
As a result, the number of phase signals in each $\Gamma_l$ is $M$. The scheme is described in Fig.9.
The second PRNG-2 choses one communication basis at each selected mapping patern.
\\

\subsubsection{Structure of phase signals as ciphertext}
The crucial point of this method is the shift permutation in the mapping and the degree of $\delta$.
A mapping pattern ${\cal{L}}_k, k=1,2,3,\dots, M$ is chosen by the random sequence of $log M$ bits from the PRNG-1
 with the secret key $K_{S1}$. 
After the selection of the mapping pattern, the second PRNG-2 with the secret key $K_{S2}$ assigns which basis should
be used to transmit the information bit.
That is, the physical phase signal is the function of two running keys and the plaintext as folllows:
\begin{equation}
 \theta_m =f(K^{R1}_i, K^{R2}_j,X ),m=\{1,2,3, \dots,2M^2\}
\end{equation}
where the running key from PRNG-1 is  $K^{R1}_l: \{l=1,2,3,\dots, M\}$, and that of PRNG-2 is $K^{R2}_j: \{j=1,2,3,\dots, M\}$.
Then, the plaintext 0 or 1 is sent by the basis selected from the communication basis of $M^2$. 
So the phase signals are determined by the combination of $K^{R1}_l$ and $K^{R2}_j$, and the plaintext $X$.

Recall that  $\Gamma_l$ is the region obtained by dividing the circumference of the original phase signal structure evenly by $2M$.
So we have $2M$ regions on the circumference of $2\pi|\alpha|$.
The phase signals of each region  $\{\Gamma_l\}, l=1,2,3,\dots M$ of upper half of the circumference
 become as follows:
\begin{eqnarray}
\{\theta(\Gamma_1)\}&=&\{\theta_1, \theta_1 +\delta, \dots \theta_1+(M-1)\delta\} \nonumber \\
\{\theta(\Gamma_2)\}&=&\{\theta_2, \theta_2 +\delta, \dots \theta_2+(M-1)\delta\} \nonumber \\
&&\vdots  \\
\{\theta(\Gamma_M)\}&=&\{\theta_M, \theta_M +\delta, \dots \theta_M+(M-1)\delta \}\nonumber
\end{eqnarray}
When one considers the data $X$, the information structure of each phase signal set has 
the following relation at each region $\Gamma_l$ of the total region $l=1,2,\dots, 2M$.
\begin{eqnarray}
\{\theta(\Gamma_1)\}&=&\{f(K^{R1}, K^{R2},X)\} \nonumber \\
\{\theta(\Gamma_2)\}&=&\{f(K^{R1}, K^{R2},X)\} \nonumber \\
&&\vdots \nonumber \\
\{\theta(\Gamma_{2M})\}&=&\{f(K^{R1}, K^{R2}, X)\}
\end{eqnarray}
Thus, the information structure is the same in each $\Gamma_l$ region.
Then, in the sense of a mathematical relationship, the above signal can be considered to map to a coherent state 
optical signal as follows:
\begin{equation}
\{\theta(\Gamma_l)\} \rightarrow |\alpha e^{i\{\theta(\Gamma_l)\}}>
\end{equation}

The decision problem for Bob and Eve becomes as follows:\\
(i) Bob can use the phase control device based on the information of PRNG1 and PRNG2, and his problem becomes 
the discrimination to the binary 
coherent states.
\begin{equation}
\{|\alpha >, |\alpha e^{i\pi} >\}
\end{equation}
This is independent of communication basis.
So he can adopt the Helstrom reeiver or the homodyne receiver.
Consequently, he can obtain the plaintext without serious error.

(ii) Eve first needs to receive the phase signal and estimate the running key sequence from that observation.
Since Eve does not know the running key information at the receiving the optical signals, her problem becomes
 the $2M^2$-ary detection problem :
\begin{equation}
\{|\alpha e^{i\theta_m}>\}, m=1,2,\dots, 2M^2
\end{equation}
So the received signal becomes 
\begin{equation}
{\hat \theta}_m =f(K^{R1}_i, K^{R2}_j,X ) + q,\quad m=\{1,2,3, \dots,2M^2\}
\end{equation}
or in the combination with OSK, it is 
\begin{equation}
{\hat \theta}_m =f(K^{R1}_i, K^{R2}_j,X_{Y00} ) + q,\quad m=\{1,2,3, \dots,2M^2\}
\end{equation}
As a result, the phase signal is the single value function with three independent variables.
Eve must estimate three independent variables from the received phase signal, which may contain errors.\\

\subsubsection{Effect region of quantum noise}
Here we stress that the structure of Eq(42) is the most important.
That is, all the cryptographic information  is the same in $2M$ block regions of Eq(42) on the circumference on the phase space.

Here, let us consider the visualization of quantum noise effect.
When the quantum noise effect is considered on the phase space, the measured value can be described 
by an analog receiver such as heterodyne receiver. In a such situation,
one has to estimate the phase signals  from the analog variable. 
The masking by quantum noise in the heterodyne receiver affects several blocks of Eq(42) close to the true phase.
That is, the masking of signal structure corresponds to the standard deviation $\sigma_{he}$ of quantum noise 
by the heterodyne receiver. 
The number of phase signals are designed under the fixed amplitude $|\alpha|$ as follows
\begin{equation}
\sigma_{he} > \Lambda \times 2M\delta=\Lambda \times 2|\theta_{j+1} - \theta_j|, \quad \Lambda =5 \sim 10
\end{equation}
This will mask some of the $\Gamma_l$ region.
Hence the information of the running keys of the first PRNG-1, the second PRNG-2 and the plaintext $X_{Y00}$ 
are completely hidden by the quantum noise (Fig.9).

\subsection{Receiver system for legitimate receiver}

The mechanism of the receiver (demodulator) of the legitimate receiver is almost the same as Fig.7.
Bob receives $2M^2$-ary phase shift keying signals. 
In order to perform proper binary signal reception for the $2M^2$-ary signal, Bob drives the phase shift device
 to the optical signal according to two PRNGs.
 
This is equivalent that Bob knows the communication basis at each slot.
In general, one could adopt Helstrom optimal quantum measurement for the binary quantum state signal: Eq(18),
 but in reality it is difficult. 
But Bob can adopt the usual homodyne receiver for the binary signal in practice.
So his error probability is given by 
\begin{equation}
{\bar P}_e^B \cong erfc(\frac{|\alpha|}{\sigma_{ho}}) \ll \frac{1}{2} 
\end{equation}
where
\begin{eqnarray}
&&erfc(y)\equiv \frac{1}{\sqrt {2\pi}}\int_y^{\infty} e^{-t^2} dt \\
&&erfc(-\infty)=1, erfc(0)=\frac{1}{2}, erfc(\infty)=0 \nonumber
\end{eqnarray}
where the amplitude of the light signal $|\alpha| \gg 1$, $\sigma^2_{ho}=1/4$ is the quantum noise effect in homodyne measurement.

\subsection{Receiver system for eavesdropper}

Since the eavesdropper does not know the secret key (and pseudo-random numbers),
 she cannot carry out the phase control on the $2M^2$-ary phase signals. 
Therefore, she has to estimate directly the optical signals as $2M^2$-ary PSK signals. 
The target quantum state signals are as follows:
\begin{equation}
\rho_m =|\alpha e^{i\theta_m}><\alpha e^{i\theta_m}|, \quad m=1,2,3,, \dots, M^2
\end{equation}
However, since the information is the same for each $\Gamma_l$ in a region consisting of $M\times \Gamma_l$, 
the quantum detection problem reduces to the identification problem of $M$ signals in a certain $\Gamma_l$.
That is, 
\begin{equation}
\rho^{\Gamma_l}_j=|\alpha e^{i\theta (\Gamma_l)_j}><\alpha e^{i\theta (\Gamma_l)_j}|, j=1,2,3,, \dots, M
\end{equation}
Thus, the error performance as the quantum optimum detection becomes as follows:
\begin{equation}
{\bar P}^E_e =\max_{\xi_j}\min_{\Pi_j}\{ 1- \sum_{j=1}^M \xi_j Tr \rho^{\Gamma_l}_j \Pi_j\}
\end{equation}
Since the signal structure is not the covariant, to find the exact solution is difficult. 
However, it may be given by Helstrom-Nakahira algorithm [32] 
which is the most general numerical analysis method in quantum detection theory.

Here, we adopt semi-classical theory to visualize the characteristics.
From the rough value of the received signal, Eve can narrow the search rigion down to the quantum noise region of 
the heterodyne receiver. Recall that all the blocks of $\Gamma_l$ are the same structure on the cryptographic information.
Thus, the signal detection problem reduces the detection for $M$ phase signals corresponding to the mapping function 
based on the running key: $K^{R_1}$ and $K^{R_2}$ in each $\Gamma_l$ (See appendix C).

Then we here can use the conventional detection rule for the discussion of the upper bound, 
assuming a priori probability is $1/M$ for each slot. 
The final result of the error performance of $M$-ary signals for the one block can be evaluated based on the Helstrom formula
 in the classical detection theory which is based on the detection probability $P_d$ between neighboring phases.
\begin{eqnarray}
{\bar P}^E_e &\cong& \frac{(M-1)}{M}(1-P_d) \nonumber \\
&=&\frac{2(M-1)}{M}erfc(\frac{\Delta}{2\sigma_{he}})
\end{eqnarray}
$\sigma^2_{he}=1$ is the quantum noise effect in heterodyne measurement.
$\Delta$ is the signal distance between neighboring phases as follows:
\begin{eqnarray}
&& \Delta \cong {\sqrt 2}|\alpha|(1-\cos\delta)^{1/2} \\
&& \delta =2\pi/M^2 \longrightarrow 0, \quad M \gg 1 
\end{eqnarray}
So, when $M=100 \sim 1000$, we have 
\begin{equation}
{\bar P}^E_e \rightarrow 1-\frac{1}{M}
\end{equation}

Thus, the channel model for each $\Gamma_l$ of the Eve's measurement process in the case of $M \gg 1$, under
 the fixed $|\alpha|, $ becomes almost uniform error for any signal and the capacity (maximum access information) can be written as follows:
\begin{eqnarray}
C_1 &\cong& \log M -\{(1-(M-1)\epsilon)\log \frac{1}{(1-(M-1)\epsilon)} \nonumber \\
&+&(M-1)\epsilon \log \frac{1}{\epsilon}\}\longrightarrow 0,\quad  M \gg 1
\end{eqnarray}
where $\epsilon =2erfc(\frac{\Delta}{2\sigma_{he}})/M$.
Thus all information of $f(K^{R1}_j, K^{R2}_j, X)$ conveyed by phase signals
are masked by quantum noise. 
As a result, at least, the unicity distance:Eq(22) can be guaranteed as follows:
\begin{eqnarray}
|K_{S1}|&\ll& n^Q_1(QNDM,K_{S1}) \le 2^{|K_{S1}|}\\
|K_{S2}|&\ll& n^Q_1(QNDM,K_{S2}) \le 2^{|K_{S2}|}
\end{eqnarray}
where, in general, $|K_{S1}|=|K_{S2}| \cong 256 \sim 1000$ which is the typical number of the conventional mathematical cipher.
This is the result of reviewing the results of the previous paper [29] from the perspective of unicity distance.

\subsection{Cryptographical perspective of QNDM}

Here we explain cryptographical properties of QNDM described above.
QNDM has the form of a product cipher in conventional cryptography, but its function is quite different.
Two mathematical ciphers are used to guarantee that the measurement results of Eve are effectively perfectly random 
when she measures the physical ciphertext.
That is, QNDM is a technique that embeds two running keys  of information in a small quantum noise region, 
which is equivalent to diffusing the quantum noise effect.

The only information from the rough phase position is the relationship between the two running keys at each $\Gamma_l$, 
but since both are completely masked by quantum noise, that information is meaningless.

Thus, in a generalized quantum stream cipher with QNDM, when Eve measures the physical ciphertexts Eq(46) 
that carry the information Eq(41), the measurement value becomes a completely random  due to quantum noise. 
Eq(55) and Eq(59) guarantee such situation.
Therefore, the ciphertext becomes a completely random number, regardless of the structure of the plaintext.
Any correlation contained in the pseudorandom numbers is made uncorrelated by quantum noise. 
Thus the encryption process becomes non injective, and the security is enhanced. 
As a result, the Shannon impossibility theorem is lifted in the strict sense.(Appendix D)\\

\section{Deliberate Signal Randomization (DSR) method by Yuen-Kumar-Nair}
In this section, we introduce the security performance of a generalized quantum stream cipher equipped with DSR to compare with QNDM.

Here, let us assume a phase shift keying (PSK) quantum stream cipher with DSR proposed by Yuen.
The DSR means as follows: 
When the true phase signal is $\theta_m=f(K^R_j,X_{Y00})$, from $\theta_m$ one randomies it to $\theta_r$
 according to a probability density $p(\theta_r|\theta_m)$. 
The range of $\theta_r$ is $\{\theta_m -\pi/2, \theta_m +\pi/2\}$[13,23].

Here we assume that the eavesdropper adopts heterodyne receiver, which is the highest performance for asynchronous
 quantum state signals. 
 In the standard quantum stream cipher, the eavesdropper receives $2M$ original phase signals to obtain  $M$ valued information
  on the running key.
The difference between signals is $\Delta_p =\pi |\alpha|/M$, Let $\sigma_{he}$ be the masking 
effect of the signal by quantum noise. 
The amount of signal masking is $\Gamma_Q=M\sigma_{he}/\pi |\alpha|$. 

When the DSR is equipped, the phase signals are spread by an amount of plus and minus $|R_p|$ from the correct signal.
$|R_p|$ is the strength of DSR that spreads quantum noise effect.
Here, the range of DSR can be set as follows.

\begin{equation}
1\le \sigma_{he} |R_p|<  \frac{1}{2}\pi |\alpha|
\end{equation}
The equivalent quantum noise of the optical heterodyne measurement is $\sigma^2_{he} =1$, and the maximum mutual information in
 the wedge approximation is [23]
\begin{equation}
C_{Hetero}\cong{\log_2 \frac{\pi|\alpha|}{2|R_p|}}
\end{equation}
Then one has Nair-Yuen formula of the generalized unicity distance for KPA as follows:
\begin{equation}
 n^Q_1 > \frac{|K|}{\log_2 \frac{\pi|\alpha|}{2|R_p|}}
\end{equation}
 If the strength of DSR is $|R_p|\cong \frac{1}{2}\pi |\alpha|$, then $C_{Hetero} \rightarrow 0$. 
 So the unicity distance Eq(22) is, at least, as follows:
\begin{equation}
 |K|\ll n^Q_1(DSR) \le 2^{|K|}
\end{equation}
This cannot be also achieved using only mathematical cipher.  

\section{Progress in experimental studies of standard and generalized quantum stream cipher}
\subsection{Experimental validation of the KCQ principle}
The aim of this research is to realize cutting-edge optical communications based on the applications of quantum effects.
The experimental research has been concerned with the development of equipment for standard quantum stream ciphers
 based on the advanced optical communication technologies 
 and the confirmation of their transmission performance in the real world, which was pioneered by Kumar and his group.
The first experiment was reported at QCMC-2002 held at MIT [33], where Shapiro was the executive chairman (Fig.10).
These activities have led to new directions in advanced optical communications.
Through those research and development,the system performance with data transmission rates of about 1 Gbps to 100 Gbps 
have been demonstrated [34$\sim$43]. Using quantum stream cipher with OSK, Tanizawa demonstrated 10,000 Km transmission 
for the application to undersea cable safety assurance [39] and P.Winzer and his group realized 160 Gbps$\sim$ 256 Gbps system [38,42].
These experiments proved the effectiveness in a real-world environment of cryptographic techniques 
that are not bound by the Shannon impossibility theorem. These studies have led to the completion of high speed technology.

\subsection{The frontier to generalization}
In order to commercialize this technology, it is necessary to realize a generalized quantum stream cipher (Quantum Enigma Cipher)
that has sufficient information-theoretically security against known plaintext attack on secret key.
To this end, experimental research has been started by Futami based on Ministry of Defense Fund.
In 2023, a generalized quantum stream cipher due to the randomization method of DSR was experimentally implemented 
by the Futami group (Fig.11) [44]. The system provides encrypted transmission at 10 Gbps speed over 400 km of optical fiber.
This is the world's first symmetric key cipher with sufficient information theoretic security on key.
However, there is no experimental study on QNDM as final target yet.\\

\begin{figure}
\centering{\includegraphics[width=7cm]{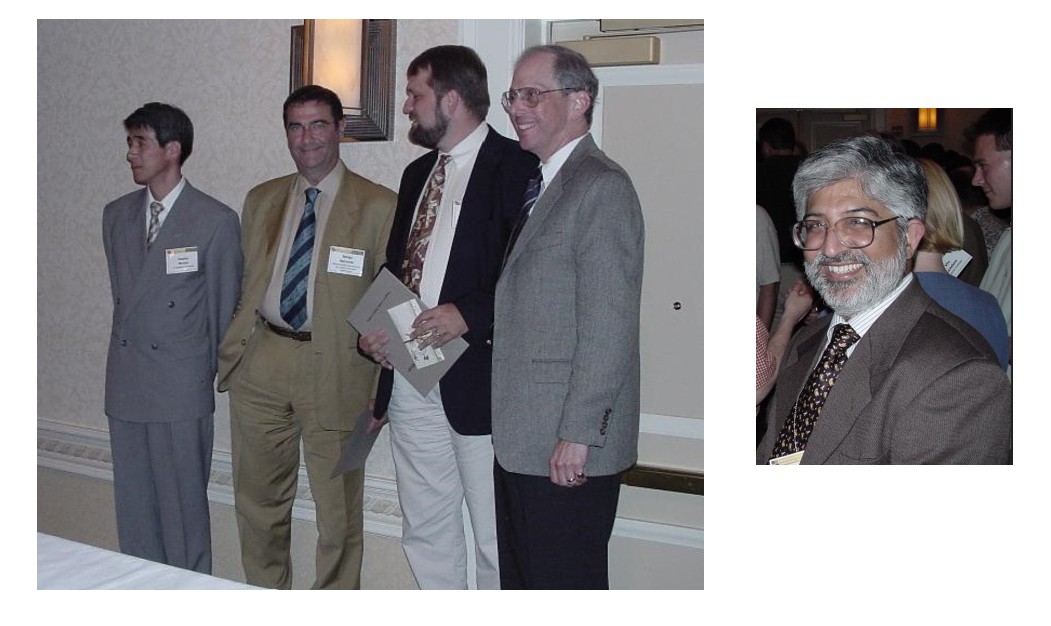}}
\caption{S.Haroche, B.Schumacher, J.H.Shapiro and P.Kumar at MIT, 2002 (by Hirota)}
\end{figure}

\begin{figure}
\centering{\includegraphics[width=7cm]{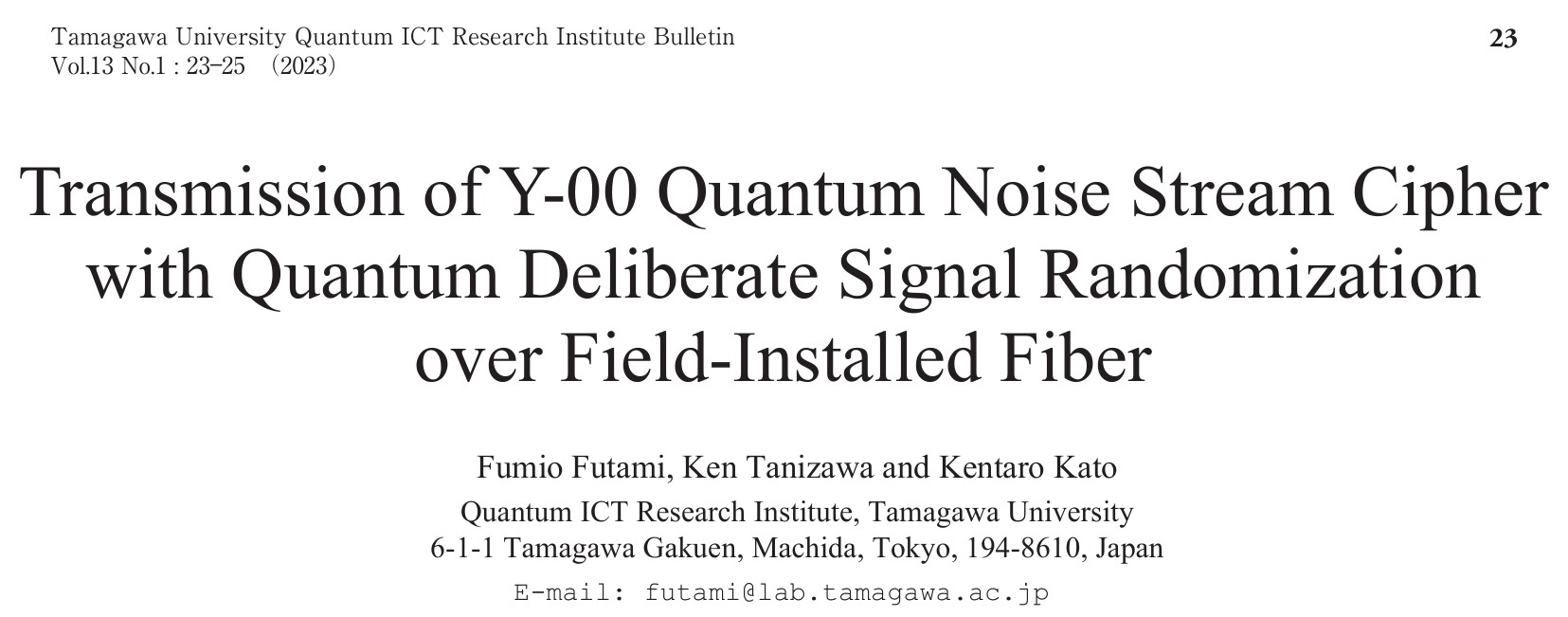}}
\caption{First demonstration of the generalized quantum stream cipher with randomization based on DSR. 
Bulletin of Quantum ICT Research Institute of Tamagawa Univ. Open access [44]}
\end{figure}

\begin{figure}
\centering{\includegraphics[width=7cm]{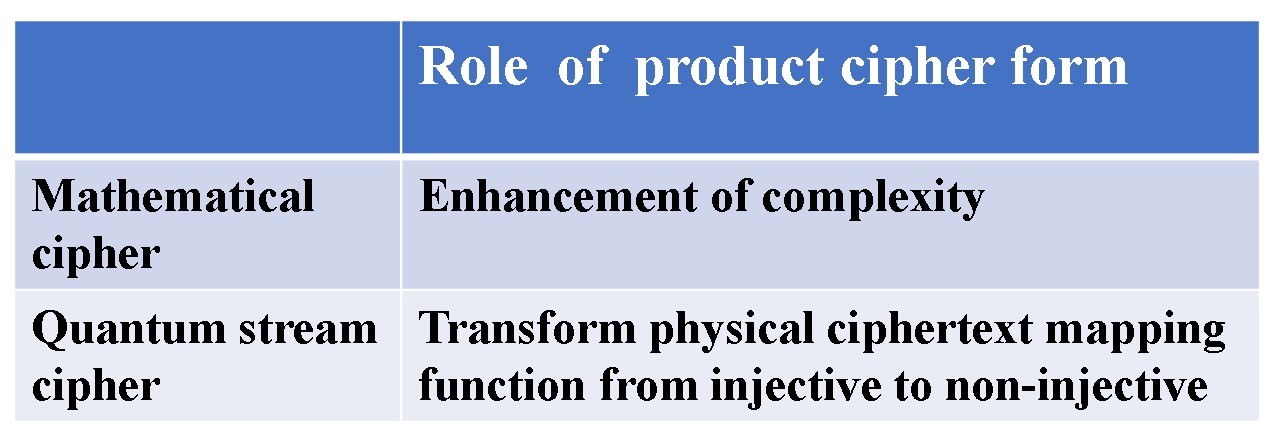}}
\caption{The role of product cipher form in quantum stream cipher and conventional mathematical cipher.}
\end{figure}

\begin{figure}
\centering{\includegraphics[width=8cm]{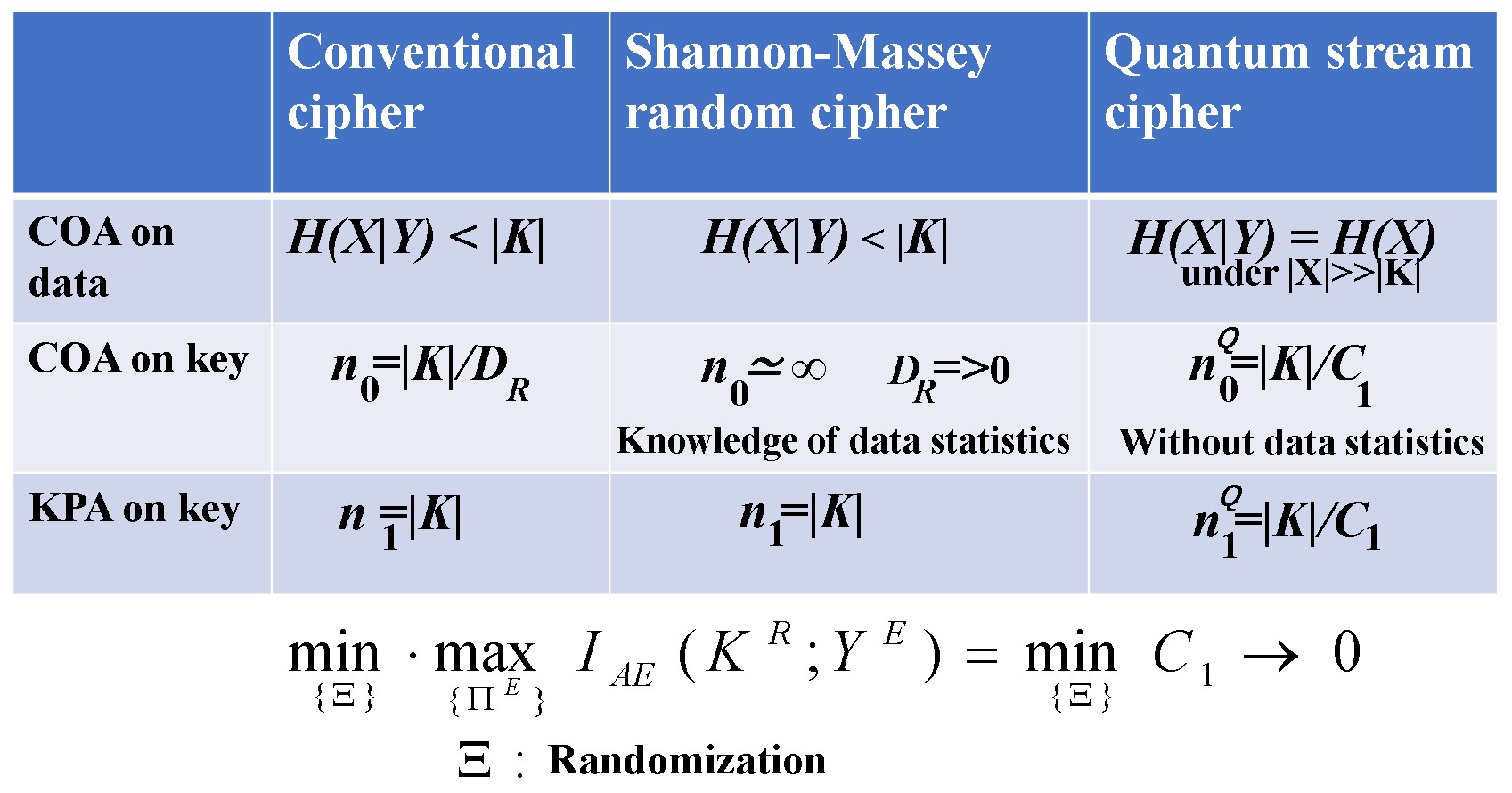}}
\caption{Security performance of the conventional cipher and the quantum stream cipher. Data means plaintext. 
$D_R$ is the redundancy of data.
 $C_1$ is the maximum mutual information of Alice-Eve measurement channel which depends on randomization of ciphertext. 
 The difference between data randomization in existing cryptography and ciphertext randomization in quantum stream cryptography is enormous.}
\end{figure}

\section{Conclusion}
Here let us recall the concept of generalized random cipher.
It is to mask by noise the ciphertext or running key sequence that the eavesdropper can obtain.
Various ideas will be able to adopt for realization of such functions, depending on the application. 
Therefore, inventing a different methodology than the one described in this series is also encouraged.
In this paper, we have explained the quantitative characteristics of the unicity distance for the randomization due to product cipher form 
(Fig.12) that extends standard quantum stream ciphers to generalized quantum stream ciphers of Type-I [13].
As a result, the QNDM amd DSR have the same performance in the sense of the unicity distance, but Bob's error performance 
is different.
In this way, it has been proven that generalized quantum stream ciphers with the additional randomizations have
 sufficient information-theoretic security on key for practical use.
 
We do not intend to claim absolutely secure cryptographic technology. Our goal is to improve the information-theoretic security on key 
 of existing mathematical cryptography by using physical phenomena. 
The advantage on the security performance of the current quantum stream cipher is shown in Fig.13.
A whole theoretical framework has not yet been completed. We will continue to work toward our final goal through 
this series following Yuen's concept, including the new wiretap channel model [45].
In addition, we aim to construct an unified theory that includes the concepts of Shapiro [46] and Lloyd [47] 
which were proposed in MIT.

\section*{Appendix}

\subsection{\textbf{Starting model of security analysis}}
When a physical quantum ciphertext is transmitted, an eavesdropper has no way of obtaining the information other than 
by performing a quantum measurement.
The goal of an eavesdropper's attack is to decrypt either the plaintext (data) or the PRNG's initial value (the secret key).
Here the security model of the basic model of standard quantum stream cipher is shown below.

In order for Eve to infer the plaintext (data) information, binary reception must be performed by regarding the transmitted signal 
as two mixed-state signals (Eq20). Eve's error characteristic at this time is given by Eq(21).
On the other hand, in order to decrypt the secret key, it is necessary to switch to a reception method that regards 
the transmitted signal as a $2M$-value pure states :Eq(17) and distinguishes between them. 
Then she collects the running key sequence. At this time, unlike ordinary encryption, 
the received value can be processed independently of the plaintext (data). 
The error characteristic at this time is given by Eq(19).

On one hand, it is impossible to deduce the plaintext from the reception of the $2M$ signal,
since the adjacent signals corresponding to plaintext ($K^R$=even, X=0; $K^R$=odd, X=1) are 
completely masked by the quantum noise as explained in the section IV.
Thus, the starting model for the security analysis becomes Fig.14.

Following Yuen's suggestion, we have been conducting research into ways to strengthen security based on this basic structure.
Specific methods are detailed below.

 \begin{figure}
\centering{\includegraphics[width=8cm]{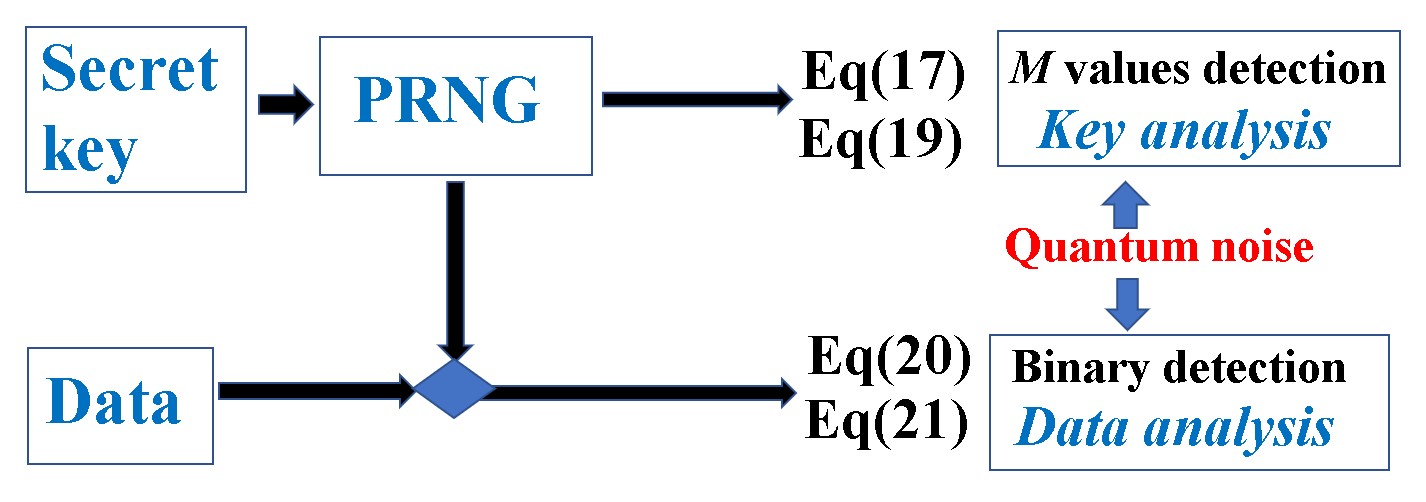}}
\caption{Starting model for security analysis without OSK and other randomization. 
For data analysis, the plaintext is hidden by quantum noise when Eve adopts the binary detection scheme.
When she adopts $2M$-ary detection, the plaintext corresponds to each adjacent signals of $2M$ signals, and 
they are hidden by quantum noise completely.
For key analysis, the running key can be measured when she adopts $M$-ary detection,
 and they are partially hidden by quantum noise.} 
\end{figure}

\subsection{\textbf{The role of OSK and QNDM}}

OSK is a method for making the mapping function that determines the physical signal ciphertext using a running key 
and plaintext ``non-injective". In addition, for direct reception of  plaintext, the OSK provides Eq(39).
Therefore, for a ciphertext-only attack on the plaintext (data), the following holds:
\begin{equation}
H(X|Y^{E_Q})=H(X)
\end{equation}
under the condition $|K| \ll |X|$. This is a perfect guarantee of security when only the data is considered (COA on data), 
but cryptography theory requires that we guarantee the security of the key.

Here we analyze in detail a security on key in a standard quantum stream cipher with OSK.
First, assume that the true plaintext sequence is leaked to an eavesdropper. 
Eve will store the measured physical ciphertext sequence containing noise.
An exhaustive search with the known plaintext sequence may be attempted for the stored signal sequence using 
a pseudorandom running key sequence to make a binary threshold decision, as would be performed by a legitimate receiver. 

On the other hand, in the case with OSK, the physical ciphertext becomes a non-injective mapping function.
Thus, even if an exhaustive search with the running key sequence to the measured signal value under the known plaintext is performed, 
a combination of the measured value and the plaintext does not give the correct information of key because of non-injective. 
Hence the known plaintext attack is reduced to a ciphertext only attack.
As a result, the unicity distance of KPA is equivalent to that of COA (See Fig.13).

Next, let us consider whether an exhaustive search of the Y-00 plaintext:$X_{Y00}$ itself is possible. 
When there is no noise, the signal value accurately indicates the running key, so even for non-injective signal values,
the Y-00 plaintext can be obtained with only the information of the signal value. 
This Y-00 plaintext $X_{Y00}$ has the form of a mathematical cipher using PRNG for the communication basis selection, 
so a known plaintext attack using the true plaintext is possible.
However, in reality, the value of the multi-valued running key is not determined within the range where the signal value
is masked by quantum noise, and the binary Y-00 plaintext sequence carried in the communication basis corresponding 
to the running key value is completely randomized by quantum noise. 
Therefore, since the ciphertext of the mathematical cipher is structured to be masked 
 completely by quantum noise, an exhaustive search against $X_{Y00}$ does not work.
 
However, in the high power laser system, the OSK alone cannot guarantee sufficient information-theoretic security on key.
 This is because the quantum noise mask can only hide part of the running key information.
So a combination of OSK and QNDM or DSR is necessary. These are the generalized quantum stream ciphers.

\subsection{\textbf{Structure of phase signals in QNDM}}
Let us describe the detailed structure of phase signals in the scheme with QNDM discussed in the section VI.
The importance is why partial masking masks all the key information.
 The phase signals carrying the encryption information belonging to each block of $M$ blocks ($\Gamma_1, \Gamma_2, \dots, \Gamma_M$) 
 on the signal phase plain have the following structure.
\begin{equation}
\theta^{\Gamma_1}_m=
\begin{cases}
\theta^{\Gamma_1}_1=\theta_1 +0 =f(K^{R_1} =1, K^{R_2} =1, X) \\
\theta^{\Gamma_1}_2=\theta_1 +\delta =f(K^{R_1} =2, K^{R_2} =2, X) \\
\vdots \\
\theta^{\Gamma_1}_M=\theta_1 +(M-1)\delta \\
=f(K^{R_1} =M, K^{R_2} =M, X) \\
\end{cases} \nonumber
\end{equation}

\begin{equation}
\theta^{\Gamma_2}_m=
\begin{cases}
\theta^{\Gamma_2}_1=\theta_2 +0 =f(K^{R_1} =1, K^{R_2} =2, X) \\
\theta^{\Gamma_2}_2=\theta_2 +\delta =f(K^{R_1} =2, K^{R_2} =3, X) \\
\vdots \\
\theta^{\Gamma_2}_M=\theta_2 +(M-1)\delta \\
=f(K^{R_1} =M, K^{R_2} =1, X) \\
\end{cases}\nonumber
\end{equation}

\begin{equation}
\theta^{\Gamma_3}_m=
\begin{cases}
\theta^{\Gamma_3}_1=\theta_3 +0 =f(K^{R_1} =1, K^{R_2} =3, X) \\
\theta^{\Gamma_3}_2=\theta_3 +\delta =f(K^{R_1} =2, K^{R_2} =4, X) \\
\vdots \\
\theta^{\Gamma_3}_M=\theta_3 +(M-1)\delta \\
=f(K^{R_1} =M, K^{R_2} =2, X) \\
\end{cases}\nonumber
\end{equation}
The above relationship is maintained up to $M$ blocks: $\theta^{\Gamma_1}_m, \theta^{\Gamma_2}_m, \dots, \theta^{\Gamma_M}_m$.

If several blocks are masked by quantum noise, the cryptographic information is completely hidden by quantum noise, 
because the cryptographic information is concentrated inside each block. 
Consequently, no information can be obtained from measuring the phase signal masked by noise.

\subsection{\textbf{KCQ vs Quantum Data Locking from the Shannon Impossibility Theorem's perspective}}
\subsubsection{\textbf{KCQ}}

The concept and its mathematical formulation of KCQ (Keyed communication in quantum noise) were proposed in 2000 
and its concrete example was disclosed in 2002 at QCMC-2002. 
The quantum stream cipher based on KCQ follows the principle of the quantum communication theory pioneered by 
Helstrom, Holevo, and Yuen.
A set of two coherent states that transmit binary classical information are used as communication basis, 
and many communication bases with different classical parameters are prepared. 
That is, one communication basis from a set of $\{|\alpha e^{\theta_j}>,|\alpha e^{\theta_j +\pi}>\}$:  $j=1,2,\dots, M$
is randomly selected using running key $j \in K_R$ from PRNG with a secret key. This corresponds to phyisical encryption. 
Then classical binary information is transmitted using this. Any modulation scheme can be used to 
randomly select the basis, but in the case of phase modulation, one can use an unitary transformation $U_{K_R}$ 
which set the communication basis based on running key.
Since the legitimate receiver knows the key, it can perform an inverse unitary transformation $U^{-1}_{K_R}$ 
with shared information. The inverse transformation transforms all bases into the starting 
communication basis $\{|\alpha>,|\alpha e^{\pi}>\}$ , 
so that Bob always receives binary PSK. Therefore, the error can be made extremely small. 
Since Eve does not know the secret key, she has to assume that the information is transmitted in all possible communication basis, 
which will result in a large error.
In other words, there will be a difference in the reception characteristics when the key is known and when it is not.
This is based on the principle of quantum communication theory that multi-ary signals have stronger non-orthogonality 
than binary signals, resulting in deterioration of discrimination performance.
This is called ADVANTAGE CREATION by the secret key.
If we limit to ciphertext-only attacks on data, Eve is forced to receive binary values, 
so the density operator for $n$ signal slot becomes as follows: 
\begin{equation}
\rho^E_X(t_1) \otimes \rho^E_X(t_2)\dots \otimes \rho^E_X(t_n), \quad X=0,1
\end{equation}
From Eq(39), Eve's measurement error at each slot becomes 1/2 for arbitrarily long data with arbitrary statistics. 
Thus the data is masked by a perfect random number.
As a result, we have 
\begin{equation}
H(X|Y) = H(X) >H(K)
\end{equation}

Here we stress that KCQ includes all methods for differentiating 
the reception capabilities of keyed and non-keyed receivers for quantum states that transmit classical information.
In addition, standard quantum stream ciphers and their generalizations, which are the realizations of KCQ, 
can be realized using existing high-speed mathematical encryption and optical modulation devices. \\

\subsubsection{\textbf{Quantum data locking}}
In 2004, D.P.DiVincenzo and his group reported a concept of quantum data locking [48].
Consider $j \in J:n$-bit plaintext as one code. Let $|j>$ be the orthogonal quantum state for one code. 
Prepare a key: $k \in K$ consisting of $m$ bits. 
The code quantum state is randomly transformed by unitary map $U^k$ using the secret key into one of 
a set of non-orthogonal state as follows:
\begin{equation}
|j,k>=U^k|j>
\end{equation}
This corresponds to an encryption. 
For $|X|=2^n$ codes with uniform distributiton, the combination of quantum state transformations using 
$|K|=2^m$ keys is $2^{n+m}$. 
The density operator of Eve without key information in this case is as follows [49]:
\begin{equation}
\rho^E =\sum_{j,k} 2^{-(n+m)}|j,k> <j,k|
\end{equation}

Let us describe the concrete structure as a cipher.  When secret key is of 1 bit, the communication basis is selected as follows:
\begin{eqnarray}
&&k=0, \quad basis : \quad \{|0 >, |1 >\} \nonumber \\
&&k=1, \quad basis :\quad \{|+ >, |- >\}
\end{eqnarray}
Then, the upper bound of the accessible information of Eve without key is given by
\begin{equation}
I_{acc}(without ``key") \le \frac{n}{2}
\end{equation}
Here, they interpret that one bit secret key allows us to encrypt $n/2$ bits.
Bob with key can return to the original orthogonal quantum state by performing the inverse operation of 
the unitary map using the key.
Bob can discriminate the orthogonal states without error, and he obatains the  information of 
Alice without degradation and he gets $n$ bits. 
They interpret it as decryption.
 
Thus, the principle is the differentiation by means of with key and without key.
Furthermore, this differentiation arises from the fact that a non-keyed receiver needs to distinguish between 
sets of non-orthogonal signals.
This results in a smaller amount of access information.
According to Holevo theory, the amount of accessible information $I_{acc}(X;Y)$ becomes smaller as the non-orthogonality 
of the signal quantum state becomes stronger.
Thus it is  another way of expressing Helstrom-Holevo-Yuen principle of the indistinguishability of non-orthogonal states.
The settings up to this point are the same as KCQ. Thus, it can be considered a form of KCQ (Fig15).

Let's return the discussion to cryptographic mechanism.
Here, the accessible information of the receiver without key can be controlled by unitary map and key length. 
That is, the unitary map corresponds to encryption in the same way as KCQ.
But they go in a different direction.
They adopt as the evaluation of security the difference between the accessible information of receiver with key and without key:
\begin{equation}
I_{acc}(with``key") -I_{acc}(without``key")
\end{equation}
where $I_{acc}$ with key is Bob's information (in general $=\log |X|$) and $I_{acc}$ without key is Eve's information.   
They use to evaluate the cryptographic feature the following function [48,50].
\begin{eqnarray}
\eta&=&\frac{H(K)}{I_{acc}(with``key") -I_{acc}(without``key")}\nonumber \\
&=& \frac{H(K)}{H(X|Y)}
\end{eqnarray}
Consequently, if the situation of $\eta < 1$ is possible, it provides the following relation like Eq(67):
\begin{equation}
H(X|Y)=\frac{1}{\eta} H(K) > H(K)
\end{equation}

Let us describe an example. 
If one requires the following upper bound
\begin{equation}
I_{acc}(without``key") < \epsilon \log |X|,\quad 0 < \epsilon < 1
\end{equation}
the required secret key entropy is [51]: 
\begin{equation}
H(K) \cong 4\log \frac{1}{\epsilon}
\end{equation}
In this case, by making $n=\log |X|$ larger, $\eta$ can be made arbitrarily small.
For example, one has the following[47].
\begin{equation}
\eta \sim \frac{\log n}{n} \ll 1, \quad n \gg 1
\end{equation}
Thus they claimed that it means the Shannon impossibility theorem is violated.
Such a mechanism is similar to a form of block cipher that masks long lengths of data with completely random short key [52].

This formalism does not include any elements that are absolutely necessary for encrypted communication for high speed data flow, 
such as a theoretical construction on the subject of time axis and its associated key requirements, communication rate (bit/sec efficiency) and 
bandwidth requirement, delay at processing and so on. Especially it requires at least 10 Gbit/sec.
Further refinements are needed to apply quantum data locking to practical network as a form of KCQ.

\begin{figure}
\centering{\includegraphics[width=8cm]{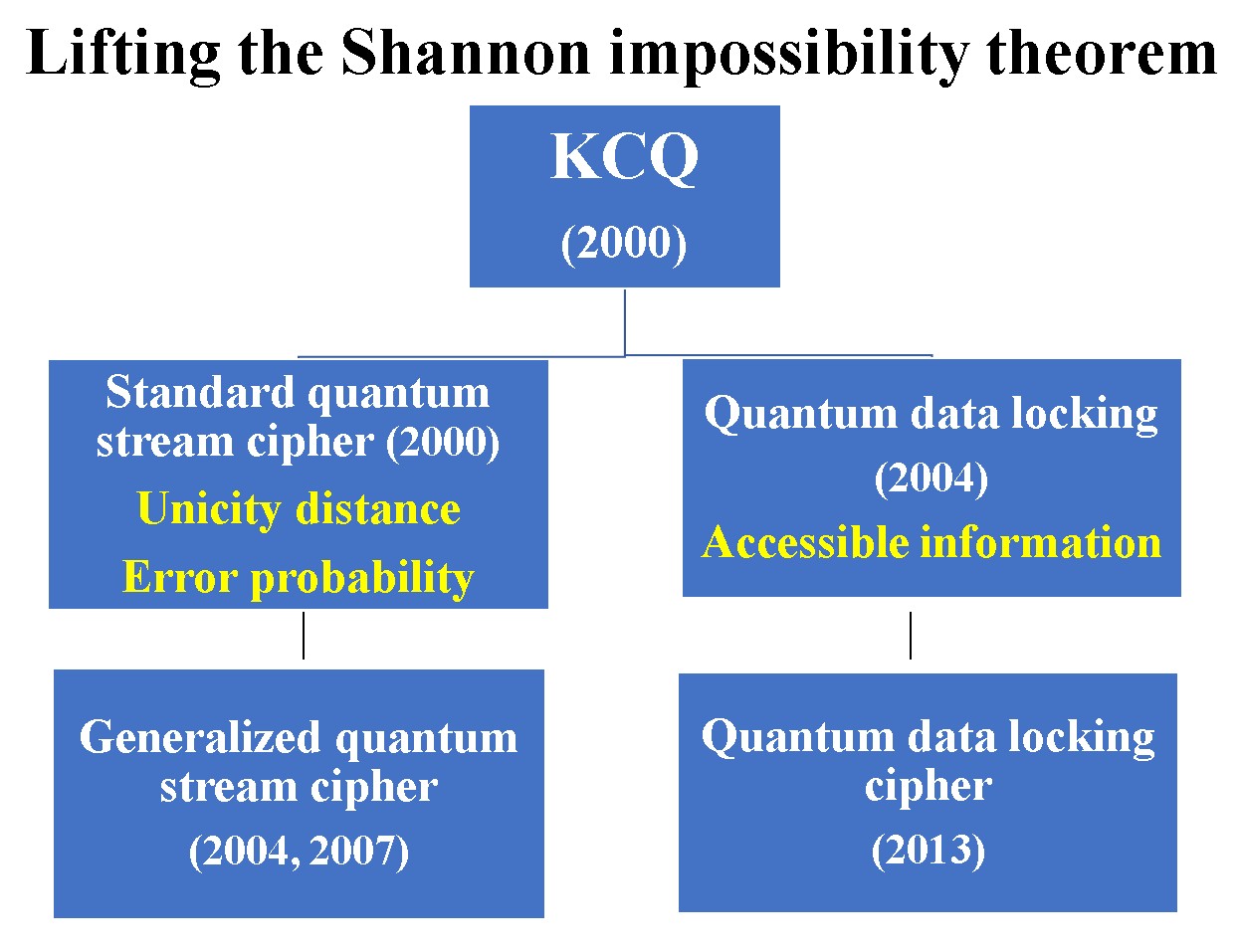}}
\caption{The origin of the lifting the Shannon impossibility theorem is KCQ (Keyed communication in quantum noise). 
As a method for achieving it, standard quantum stream cipher was proposed in 2000 and experimentally demonstrated in 2002 [33][34]. 
Generalized version was given in 2004[23] and 2007 [29].
In 2004, a method called quantum data lock was proposed from a different perspective [48], 
and in 2013 it began to be discussed as an encryption method [47][49]. 
Security criteria is unicity distance and accessible information, respectively.}
\end{figure}

\subsection{\textbf{Some remarks on Yuen's achievement}}
Here we introduce Yuen's contributions to quantum information science. Yuen believed that by developing quantum 
communication theory, 
it would be possible to further improve the possibilities of optical communication and optical signal processing. 
Since the limits of optical communication are limited by quantum noise, 
the challenge was to pursue technology to minimize its effects. R.S.Kennedy and S.Dolinar were
 in charge of the technology to realize the Helstrom limit predicted by C.W.Helstrom, 
 while Yuen proposed the concept of the two-photon coherent state to control the quantum state of light [53]
 and developed its role in communication theory in collaboration with J.H.Shapiro[54,55] and V.W.Chan [56]. 
  In Japan, our group proceeded its development [57,58].
 The results have now grown into a major field as the research and development of the squeezed state. 
G.J.Milburn gives a fair overview of Yuen's contributions to physics [59], and his contributions to communications
 and signal processing are covered in the author's book [60].

Finally, he showed interest in research using quantum noise as the positive role, which led to 
the concept of generalized random cipher that is introduced in this paper. 
In our seminar in 1999, he explained that the Shannon impossible theorem is not essential. 
Then, in order to develop a methodology for breaking the Shannon limit, we started to provide our numerical analysis 
of the detection characteristics for multivalues coherent state signals to him. 
In 2002, the Northwestern University group released the results of their research, including a principle experiment, 
to the public [33].
Based on his ideas, new directions for cutting-edge optical communication technology are emerging, 
although they are still in development.

On the other hand, Yuen's thoughts on constructing the most rigorous theory of quantum key distribution are 
detailed in his paper [61], 
which is the invited paper from Editor D.Abbott of IEEE Access.
  Please refer to the website of Northwestern University for furthermore details on his achievement.

In his research, he often presented rough sketches of his ideas, which led many researchers to misunderstand.
However, thanks to the efforts of his friends, his new proposals and theories have been proven to be true. 
I believe that his work will continue to be reevaluated.

 \section*{Acknowledgements}
I am grateful to M.Sohma, K.Kato for the discussions I had with them. I would like to express my gratitude to  
F.Futami, T.Usuda, K.Nakahira and K.Tanizawa for their activities.

\end{document}